\documentclass[aps,pre,reprint,tightenlines,twocolumn,showpacs,showkeys]{revtex4}
\usepackage[dvips]{graphicx}
\usepackage{hyperref,breakurl,amsmath,amssymb,amsbsy,tipa}

\newcommand{\ie}{\emph{i.e.}, }
\newcommand{\eg}{\emph{e.g.}, }
\newcommand{\ii}{\textrm{i}}
\newcommand{\ee}{\textrm{e}}
\newcommand{\dd}{\textrm{d}}
\newcommand{\DD}{\textrm{D}}
\newcommand{\re}{\textrm{Re}}
\newcommand{\im}{\textrm{Im}}
\newcommand{\Tr}{\textrm{Tr}}
\newcommand{\bTr}{\textrm{bTr}}
\newcommand{\erfc}{\textrm{erfc}}
\newcommand{\Id}{\mathbf{1}}
\newcommand{\Zero}{\mathbf{0}}
\newcommand{\la}{\left\langle}
\newcommand{\ra}{\right\rangle}

\begin{document}

\title{Summing free unitary random matrices}

\author{Andrzej \surname{Jarosz}}
\email{jedrekjarosz@gmail.com}
\affiliation{The Henryk Niewodnicza\'{n}ski Institute of Nuclear Physics, Polish Academy of Sciences, Radzikowskiego 152, 31--342 Krak\'{o}w, Poland}

\topmargin=0cm
\allowdisplaybreaks[4]

\begin{abstract}
I use quaternion free probability calculus --- an extension of free probability to non--Hermitian matrices (which is introduced in a succinct but self--contained way) --- to derive in the large--size limit the mean densities of the eigenvalues and singular values of sums of independent unitary random matrices, weighted by complex numbers. In the case of CUE summands, I write them in terms of two ``master equations,'' which I then solve and numerically test in four specific cases. I conjecture a finite--size extension of these results, exploiting the complementary error function. I prove a central limit theorem, and its first sub--leading correction, for independent identically--distributed zero--drift unitary random matrices.
\end{abstract}


\maketitle


\section{Introduction}
\label{s:Introduction}


\subsection{Model}
\label{ss:Model}


\subsubsection{Definition of the model}
\label{sss:DefinitionOfTheModel}

The main objective of this paper is to begin investigating the following non--Hermitian random matrix model,
\begin{equation}\label{eq:WDefinition}
\mathbf{W} \equiv \mathbf{S} \mathbf{P} .
\end{equation}
Here,
\begin{equation}\label{eq:SDefinition}
\mathbf{S} \equiv w_{1} \mathbf{U}_{1} + w_{2} \mathbf{U}_{2} + \ldots + w_{L} \mathbf{U}_{L} ,
\end{equation}
is a sum of $L \geq 2$ independent unitary random matrices of dimensions $N \times N$, weighted by some arbitrary complex numbers \smash{$w_{l}$}, $l = 1 , 2 , \ldots , L$. Everywhere, except subsection~\ref{ss:CentralLimitTheorem}, the \smash{$\mathbf{U}_{l}$}'s will belong to the simplest circular unitary ensemble (CUE). Moreover,
\begin{equation}\label{eq:PDefinition}
\mathbf{P} \equiv \mathbf{A}_{1} \mathbf{A}_{2} \ldots \mathbf{A}_{K} ,
\end{equation}
is a product of $K \geq 1$ complex random matrices, where \smash{$\mathbf{A}_{k}$}, $k = 1 , 2 , \ldots , K$, is rectangular of dimensions \smash{$N_{k} \times N_{k + 1}$} (hence, $\mathbf{P}$ has dimensions \smash{$N_{1} \times N_{K + 1}$}, and there must be \smash{$N = N_{1}$}; the same are the dimensions of $\mathbf{W}$), and where all the real and imaginary parts of the matrix elements of the \smash{$\mathbf{A}_{k}$}'s are independent random numbers with the Gaussian distribution of zero mean, \ie concisely,
\begin{equation}\label{eq:RectangularGGMeasure}
\textrm{JPDF} \left( \mathbf{A}_{k} \right) \propto \exp \left( - \frac{\sqrt{N_{k} N_{k + 1}}}{\sigma_{k}^{2}} \Tr \left( \mathbf{A}_{k}^{\dagger} \mathbf{A}_{k} \right) \right) ,
\end{equation}
where the \smash{$\sigma_{k}$}'s are real positive parameters (which set the respective variances to be \smash{$\sigma_{k}^{2} / ( 2 ( N_{k} N_{k + 1} )^{1 / 2} )$}). Finally, any matrix entry of any \smash{$\mathbf{U}_{l}$} is statistically independent from any entry of any \smash{$\mathbf{A}_{k}$}.

In this article, I will consider only the part $\mathbf{S}$ (\ref{eq:SDefinition}), \ie choose $K = 0$. The more general model $\mathbf{W}$ (\ref{eq:WDefinition}) will be left for a separate paper.


\subsubsection{Thermodynamic limit}
\label{sss:ThermodynamicLimit}

The tools I apply --- quaternion free probability, with its quaternion addition law (\ref{eq:QuaternionAdditionLaw})~\cite{JaroszNowak2004,JaroszNowak2006}, \ie an extension of the standard free probability addition law (\ref{eq:HermitianAdditionLaw})~\cite{VoiculescuDykemaNica1992,Speicher1994} into the non--Hermitian realm --- will allow to handle the above model only in the ``thermodynamic limit,''
\begin{equation}
\begin{split}\label{eq:ThermodynamicLimit}
&N = N_{1} , N_{2} , \ldots , N_{K + 1} \to \infty ,\\
&R_{k} \equiv \frac{N_{k}}{N_{K + 1}} \textrm{ = finite,}
\end{split}
\end{equation}
where the $K$ finite parameters \smash{$R_{k}$} are called the ``rectangularity ratios.''

However, in subsection~\ref{ss:FiniteSizeEffects}, I will conjecture a finite--size modification of the results for the model $\mathbf{S}$ (with CUE's as the summands), featuring a simple form--factor (\ref{eq:FiniteSizeBorderlineFactor}), which performs very well when numerically tested. The same form--factor should work for $\mathbf{W}$.


\subsubsection{Mean densities of the eigenvalues and singular values}
\label{sss:MeanDensitiesOfTheEigenvaluesAndSingularValues}

I will be interested in the two simplest statistical properties of the above model:
\begin{itemize}
\item The mean density of the eigenvalues (``mean spectral density'') of $\mathbf{W}$ (in this paper, only of $\mathbf{S}$),
    \begin{equation}\label{eq:NonHermitianMeanSpectralDensityDefinition}
    \rho_{\mathbf{W}} ( z , z^{*} ) \equiv \frac{1}{N} \sum_{i = 1}^{N} \la \delta^{( 2 )} \left( z - \lambda_{i} \right) \ra .
    \end{equation}
    There must be \smash{$N_{K + 1} = N$} (\ie \smash{$R_{1} = 1$}) for $\mathbf{W}$ to be a square matrix. Also, the averaging is performed with respect to the probability measure of $\mathbf{W}$, and the complex Dirac delta is used because the eigenvalues \smash{$\lambda_{i}$} are generically complex.
\item The mean density of the singular values, defined as the (real and non--negative) eigenvalues of the Hermitian random matrix \smash{$\mathbf{H} \equiv \mathbf{W}^{\dagger} \mathbf{W}$},
    \begin{equation}\label{eq:HermitianMeanSpectralDensityDefinition}
    \rho_{\mathbf{H}} ( x ) \equiv \frac{1}{N_{K + 1}} \sum_{i = 1}^{N_{K + 1}} \la \delta \left( x - \mu_{i} \right) \ra .
    \end{equation}
    (According to another terminology, these would be the singular values squared.) In this case \smash{$R_{1}$} can be arbitrary, and $\mathbf{H}$ has dimensions \smash{$N_{K + 1} \times N_{K + 1}$}. The real Dirac delta is exploited since the singular values \smash{$\mu_{i}$} are real.
\end{itemize}


\subsection{Motivation}
\label{ss:Motivation}


\subsubsection{Study of non--Hermitian random matrices}
\label{sss:StudyOfNonHermitianRandomMatrices}

The model $\mathbf{W}$ (\ref{eq:WDefinition}) is interesting from the mathematical point of view, since it is non--Hermitian --- and such random matrices have beautiful mathematical structure, more involved than Hermitian ones, plus multiple physical applications, ranging from finances and biology to quantum physics (for a review, consult \eg~\cite{KhoruzhenkoSommers2009}). In particular, the problems of summing (\eg~\cite{Stephanov1996,FeinbergZee1997-01,FeinbergZee1997-02,JanikNowakPappWambachZahed1997,JanikNowakPappZahed1997-01,JanikNowakPappZahed1997-02,HaagerupLarsen2000,GorlichJarosz2004,Rogers2010,JaroszNowak2004,JaroszNowak2006}) and multiplying (\eg~\cite{GredeskulFreilikher1990,CrisantiPaladinVulpiani1993,Beenakker1997,Caswell2000,JacksonLautrupJohansenNielsen2002,JanikWieczorek2003,GudowskaNowakJanikJurkiewiczNowak20032005,TulinoVerdu2004,NarayananNeuberger2007,BanicaBelinschiCapitaineCollins2007,BlaizotNowak2008,LohmayerNeubergerWettig2008,BenaychGeorges2008,KanzieperSingh2010,BurdaJanikWaclaw2010,BurdaJaroszLivanNowakSwiech20102011,Jarosz2010-01,PensonZyczkowski2011,Rogers2010}) non--Hermitian random matrices have been drawing considerable attention --- and the model $\mathbf{W}$ includes both these operations.


\subsubsection{Applications to quantum entanglement}
\label{sss:ApplicationsToQuantumEntanglement}

The model $\mathbf{W}$ (\ref{eq:WDefinition}) arises in the theory of random quantum states (see the textbook~\cite{BengtssonZyczkowski2006} and~\cite{ZyczkowskiPensonNechitaCollins2010,SEMZyczkowski2010} for review; I base this introduction on these latter works). Such objects are used for instance to describe states of a quantum system affected by noise, \ie complicated interactions with an environment which can be regarded as random. Also, if one looks for generic properties of a complicated quantum state, one may assume it random. A random quantum state is defined by specifying a probability measure in the space of density matrices $\boldsymbol{\rho}$, \ie Hermitian, weakly positive--definite (\ie with non--negative eigenvalues) and normalized (\ie $\Tr \boldsymbol{\rho} = 1$) matrices. One way to do this is to take any rectangular matrix model $\mathbf{X}$, and then \smash{$\boldsymbol{\rho} \equiv \mathbf{X} \mathbf{X}^{\dagger} / \Tr ( \mathbf{X} \mathbf{X}^{\dagger} )$} is a proper random quantum density matrix.

More precisely, if one considers a bi--partite system consisting of a principal system $\mathcal{A}$ of size \smash{$N_{1}$} and an environment $\mathcal{B}$ of size \smash{$N_{2}$}, one may form a pure state as a linear combination of the product basis,
\begin{equation}\label{eq:QuantumEntanglementDerivation1}
| \psi \rangle \equiv \sum_{i = 1}^{N_{1}} \sum_{j = 1}^{N_{2}} X_{i j} | i \rangle_{\mathcal{A}} \otimes | j \rangle_{\mathcal{B}} .
\end{equation}
Now, a mixed state on the principal system is obtained by taking the ``partial trace'' over the environment,
\begin{equation}\label{eq:QuantumEntanglementDerivation2}
\boldsymbol{\rho} \equiv \frac{\Tr_{\mathcal{B}} | \psi \rangle \langle \psi |}{\langle \psi | \psi \rangle} = \frac{\mathbf{X} \mathbf{X}^{\dagger}}{\Tr \left( \mathbf{X} \mathbf{X}^{\dagger} \right)} .
\end{equation}
Then, different probability distributions of the pure states (\ie of $\mathbf{X}$) lead to different ensembles of quantum density matrices.

The model $\mathbf{S}$ (\ref{eq:SDefinition}) thus appears in the following application of the above program: Consider a bi--partite system $( \mathcal{A} , \mathcal{B} )$ of size $N \times N$ (two ``quNits''), and a maximally entangled state (the ``generalized Bell state'') on it,
\begin{equation}\label{eq:QuantumEntanglementDerivation3}
| \Psi^{+}_{\mathcal{A} \mathcal{B}} \rangle \equiv \frac{1}{\sqrt{N}} \sum_{i = 1}^{N} | i \rangle_{\mathcal{A}} \otimes | i \rangle_{\mathcal{B}} .
\end{equation}
(A ``typical'' random state is entangled because there are much more entangled states than separable ones; the latter form a set of measure zero in the set of all states.) Now, perform on this state $L$ independent random local unitary transformations in the principal system $\mathcal{A}$,
\begin{equation}\label{eq:QuantumEntanglementDerivation4}
| \psi_{l} \rangle \equiv \left( \mathbf{U}_{l} \otimes \mathbf{1}_{N} \right) | \Psi^{+}_{\mathcal{A} \mathcal{B}} \rangle ;
\end{equation}
the resulting states remain maximally entangled. Form a probability mixture of these $L$ states, \ie a superposition with coefficients \smash{$w_{l} \in [ 0 , 1 ]$} such that \smash{$\sum_{l = 1}^{L} w_{l} = 1$}, \ie
\begin{equation}\label{eq:QuantumEntanglementDerivation5}
| \psi \rangle \equiv \sum_{l = 1}^{L} w_{l} | \psi_{l} \rangle = \left( \mathbf{S} \otimes \mathbf{1}_{N} \right) | \Psi^{+} \rangle .
\end{equation}
Finally, take the normalized partial trace over the environment $\mathcal{B}$, which leads to the random mixed state (\ref{eq:QuantumEntanglementDerivation2}) with $\mathbf{X} = \mathbf{S}$.

The normalization, \smash{$\Tr ( \mathbf{S} \mathbf{S}^{\dagger} ) = N ( \sum_{l = 1}^{L} | w_{l} |^{2} ) + \ldots$}, where the dots are \smash{$2 \sum_{l < l^{\prime}} \re ( w_{l} w_{l^{\prime}}^{*} \Tr ( \mathbf{U}_{l} \mathbf{U}_{l^{\prime}}^{\dagger} ) ) \ll N$}, if only the spectrum of the \smash{$\mathbf{U}_{l}$}'s remains finite in the thermodynamic limit (\ref{eq:ThermodynamicLimit}). Hence, it is enough to study the model \smash{$\mathbf{S} \mathbf{S}^{\dagger}$}, or practically equivalently (modulo zero modes), \smash{$\mathbf{S}^{\dagger} \mathbf{S}$}, \ie the singular values of $\mathbf{S}$.

To my knowledge, the mean density of the singular values of $\mathbf{S}$ has been so far known only for all the \smash{$w_{l}$}'s equal, which is the ``Kesten distribution''~\cite{Kesten1959}; the corresponding mean spectral density is also known~\cite{HaagerupLarsen2000,GorlichJarosz2004}.

The model $\mathbf{P}$ (\ref{eq:PDefinition}) appears in another setup: Consider a system consisting of an even number $2 K$ of subsystems, with the following sizes,
\begin{equation}\label{eq:QuantumEntanglementDerivation6}
\underbrace{\mathcal{A}_{1} ,}_{\textrm{size } N_{1}} \underbrace{\mathcal{A}_{2} , \mathcal{A}_{3}}_{\textrm{size } N_{2}} , \ldots , \underbrace{\mathcal{A}_{2 K - 2} , \mathcal{A}_{2 K - 1}}_{\textrm{size } N_{K}} , \underbrace{\mathcal{A}_{2 K} .}_{\textrm{size } N_{K + 1}}
\end{equation}
Consider an arbitrary product state \smash{$| \psi_{0} \rangle \equiv | 0 \rangle_{\mathcal{A}_{1}} \otimes | 0 \rangle_{\mathcal{A}_{2}} \otimes \ldots \otimes | 0 \rangle_{\mathcal{A}_{2 K}}$}. Now, form a random pure state by performing on \smash{$| \psi_{0} \rangle$} independent random local unitary transformations acting on the following pairs of the subsystems,
\begin{equation}\label{eq:QuantumEntanglementDerivation7}
| \psi \rangle \equiv \left( \mathcal{U}_{\mathcal{A}_{1} \mathcal{A}_{2}} \otimes \mathcal{U}_{\mathcal{A}_{3} \mathcal{A}_{4}} \otimes \ldots \otimes \mathcal{U}_{\mathcal{A}_{2 K - 1} \mathcal{A}_{2 K}} \right) | \psi_{0} \rangle .
\end{equation}
By definition, the result is a product state with respect to this latter pairing, \ie it can be expanded in the product basis as
\begin{equation}
\begin{split}\label{eq:QuantumEntanglementDerivation8}
| \psi \rangle = &\sum_{i_{1} = 1}^{N_{1}} \sum_{i_{2} , i_{2}^{\prime} = 1}^{N_{2}} \ldots \sum_{i_{K} , i_{K}^{\prime} = 1}^{N_{K}} \sum_{i_{K + 1} = 1}^{N_{K + 1}}\\
&[ \mathbf{A}_{1} ]_{i_{1} i_{2}} [ \mathbf{A}_{2} ]_{i_{2}^{\prime} i_{3}} \ldots [ \mathbf{A}_{K - 1} ]_{i_{K - 1}^{\prime} i_{K}} [ \mathbf{A}_{K} ]_{i_{K} i_{K + 1}} \cdot\\
&\cdot | i_{1} \rangle_{\mathcal{A}_{1}} \otimes | i_{2} \rangle_{\mathcal{A}_{2}} \otimes | i_{2}^{\prime} \rangle_{\mathcal{A}_{3}} \otimes \ldots \otimes | i_{K + 1} \rangle_{\mathcal{A}_{2 K}} ,
\end{split}
\end{equation}
where the coefficients are collected into $K$ matrices \smash{$\mathbf{A}_{k}$} of rectangular dimensions \smash{$N_{k} \times N_{k + 1}$}. In the simplest case, they may be assumed Gaussian (\ref{eq:RectangularGGMeasure}). Consider further the maximally entangled states (\ref{eq:QuantumEntanglementDerivation3}) on the pairs \smash{$( \mathcal{A}_{2} , \mathcal{A}_{3} )$}, \ldots, \smash{$( \mathcal{A}_{2 K - 2} , \mathcal{A}_{2 K - 1} )$}, and perform the projective measurement of $| \psi \rangle$ onto the product of these Bell states,
\begin{equation}\label{eq:QuantumEntanglementDerivation9}
\mathcal{P} \equiv \mathbf{1}_{\mathcal{A}_{1}} \otimes \left( \bigotimes_{k = 2}^{K} | \Psi^{+}_{\mathcal{A}_{2 k - 2} \mathcal{A}_{2 k - 1}} \rangle \langle \Psi^{+}_{\mathcal{A}_{2 k - 2} \mathcal{A}_{2 k - 1}} | \right) \otimes \mathbf{1}_{\mathcal{A}_{2 K}} ,
\end{equation}
which yields a random pure state describing the remaining two subsystems, \smash{$\mathcal{A}_{1}$} and \smash{$\mathcal{A}_{2 K}$},
\begin{equation}
\begin{split}\label{eq:QuantumEntanglementDerivation10}
| \phi \rangle &\equiv \mathcal{P} | \psi \rangle \propto\\
&\propto \sum_{i_{1} = 1}^{N_{1}} \sum_{i_{K + 1} = 1}^{N_{K + 1}} [ \mathbf{P} ]_{i_{1} i_{K + 1}} | i_{1} \rangle_{\mathcal{A}_{1}} \otimes | i_{K + 1} \rangle_{\mathcal{A}_{2 K}} .
\end{split}
\end{equation}
As a final step, take the normalized partial trace over \smash{$\mathcal{A}_{2 K}$}, thus obtaining the mixed state on \smash{$\mathcal{A}_{1}$}, (\ref{eq:QuantumEntanglementDerivation2}) with $\mathbf{X} = \mathbf{P}$, whose statistical properties are directly related to the singular values of $\mathbf{P}$.

The mean density of the singular values of $\mathbf{P}$, with the assumption of all the sizes \smash{$N_{k}$} equal (\ie all \smash{$R_{k} = 1$} (\ref{eq:ThermodynamicLimit})), has been known as the ``Fuss--Catalan distribution''~\cite{BanicaBelinschiCapitaineCollins2007,PensonZyczkowski2011}; the corresponding mean spectral density has been found in~\cite{BurdaJanikWaclaw2010}. For arbitrary \smash{$R_{k}$}'s, after a primary work on the $K = 2$ case~\cite{KanzieperSingh2010}, both mean densities have been derived in~\cite{BurdaJaroszLivanNowakSwiech20102011}.

The model $\mathbf{W}$ (\ref{eq:WDefinition}) arises when a combination of both the above procedures is applied, \ie one takes a probability mixture of $L$ random pure states defined on the $2 K$ subsystems, performs the projective measurement on the product basis of the $( K - 1 )$ maximally entangled states, and takes the partial trace over the last subsystem.

According to my knowledge, the only known result concerns the mean density of the singular values of $\mathbf{W}$ for $L = 2$, with equal weights, and $K = 1$, which is the so--called ``Bures distribution''~\cite{Bures1969,SommersZyczkowski2004}. Therefore, generalization to arbitrary $L$, $K$, weights, as well as considering the eigenvalues --- comprises an important research program, which will be accomplished in this and a forthcoming publication.


\subsubsection{Applications to random walks on regular trees}
\label{sss:ApplicationsToRandomWalksOnRegularTrees}

I will just mention that the model $\mathbf{S}$ (\ref{eq:SDefinition}), in the CUE case, finds applications to random walks on $L$--regular trees~\cite{Kesten1959}.


\subsection{Plan of the paper}
\label{ss:PlanOfThePaper}

\begin{description}
\item[Section~\ref{s:QuaternionFreeProbability}:] ``Quaternion free probability'': I introduce the fundamental notions (Green function, $M$--transform, mean spectral density) of both Hermitian and non--Hermitian random matrix theory, in the latter case especially mentioning unitary matrices (\ref{ss:HermitianAndNonHermitianGreenFunctions}). I briefly describe free probability theory of Voiculescu--Speicher and the addition algorithm, both in the Hermitian and non--Hermitian setting, the latter being the quaternion formalism, which is a crucial tool for this work (\ref{ss:QuaternionFreeProbabilityInANutshell}). I show how it is possible to reduce (``rational Hermitization'') the quaternion Green function for unitary matrices to the standard Green function (\ref{ss:RationalHermitizationProcedure}).
\item[Section~\ref{s:EIGSummingFreeUnitaryRandomMatrices}:] ``Summing free unitary random matrices --- the eigenvalues'': I employ quaternion free probability to the challenge of computing in the thermodynamic limit the mean spectral density of a weighted sum of free CUE random matrices; besides a general (``master'') equation, five specific examples are considered (\ref{ss:EIGSummingFreeCUERandomMatrices}). The central limit theorem for free identically--distributed zero--drift unitary random matrices, and its sub--leading term, are proven (\ref{ss:CentralLimitTheorem}). I introduce and verify numerically a modification of the mean spectral densities from subsection~\ref{ss:EIGSummingFreeCUERandomMatrices} which should be valid for finite matrix dimensions and which uses the complementary error function; I make a conjecture of its even broader application (\ref{ss:FiniteSizeEffects}).
\item[Section~\ref{s:SVSummingFreeUnitaryRandomMatrices}:] ``Summing free unitary random matrices --- the singular values'': I recall a conjecture relating the mean spectrum of the eigenvalues, when it is rotationally--symmetric around zero, and of the corresponding singular values (\ref{ss:ConjectureAboutRotationallySymmetricSpectra}). I exploit this hypothesis to find the master equation for the singular values, in the thermodynamic limit, of a weighted sum of free CUE random matrices; I also analyze the five examples from subsection~\ref{ss:EIGSummingFreeCUERandomMatrices} in this context (\ref{ss:SVSummingFreeCUERandomMatrices}).
\item[Section~\ref{s:Conclusions}:] ``Conclusions'': I summarize the results of the paper (\ref{ss:Summary}), and pose some unsolved problems (\ref{ss:OpenProblems}).
\end{description}


\section{Quaternion free probability}
\label{s:QuaternionFreeProbability}


\subsection{Hermitian and non--Hermitian Green functions}
\label{ss:HermitianAndNonHermitianGreenFunctions}


\subsubsection{Hermitian Green function}
\label{sss:HermitianGreenFunction}

\begin{figure}[t]
\includegraphics[width=\columnwidth]{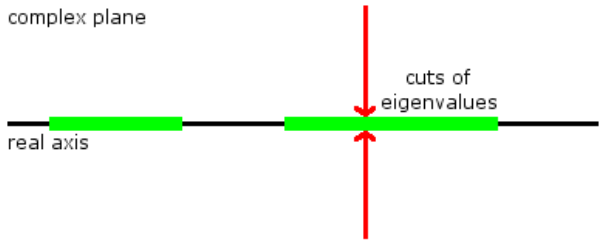}
\caption{For Hermitian random matrices, the eigenvalues are real, and one needs a complex Green function of a complex variable.}
\label{fig:EigenvaluesReal}
\end{figure}

``Quaternion free probability'' is a version of Voiculescu's free probability calculus designed to handle non--Hermitian random matrices. To introduce it, let us however begin in the Hermitian realm. The ``mean spectral density'' of an $N \times N$ Hermitian random matrix $\mathbf{H}$ (\ref{eq:HermitianMeanSpectralDensityDefinition}) is conveniently encoded in terms of the ``holomorphic Green function,''
\begin{equation}\label{eq:HolomorphicGreenFunctionDefinition}
G_{\mathbf{H}} ( z ) \equiv \frac{1}{N} \Tr \la ( z \Id_{N} - \mathbf{H} )^{- 1} \ra = \frac{1}{N} \sum_{i = 1}^{N} \la \frac{1}{z - \lambda_{i}} \ra .
\end{equation}
This is a meromorphic function, with poles coinciding with the mean (real) spectrum; in the large--$N$ limit, these poles coalesce into continuous intervals on the real axis, and this Green function turns into a holomorphic function on the whole complex plane except these cuts. The mean spectral density is retrieved from the holomorphic Green function taken in the vicinity of these cuts, through
\begin{equation}\label{eq:MeanSpectralDensityFromHolomorphicGreenFunction}
\rho_{\mathbf{H}} ( \lambda ) = - \frac{1}{2 \pi \ii} \lim_{\epsilon \to 0^{+}} \left( G_{\mathbf{H}} ( \lambda + \ii \epsilon ) - G_{\mathbf{H}} ( \lambda - \ii \epsilon ) \right) ,
\end{equation}
as follows from the representation of the real Dirac delta, \smash{$\delta ( \lambda ) = - \frac{1}{2 \pi \ii} \lim_{\epsilon \to 0^{+}} ( \frac{1}{\lambda + \ii \epsilon} - \frac{1}{\lambda - \ii \epsilon} )$}; this is illustrated in figure~\ref{fig:EigenvaluesReal}. Remark that alternatively to the Green function, one often prefers the ``holomorphic $M$--transform,''
\begin{equation}\label{eq:HolomorphicMTransformDefinition}
M_{\mathbf{H}} ( z ) \equiv z G_{\mathbf{H}} ( z ) - 1 .
\end{equation}


\subsubsection{Unitary Green function}
\label{sss:UnitaryGreenFunction}

For unitary matrices $\mathbf{U}$, the spectrum is not real, but is still one--dimensional (belongs to the centered unit circle $C ( 0 , 1 )$, \ie \smash{$\lambda_{i} = \ee^{\ii \theta_{i}}$}, \smash{$\theta_{i} \in [ 0 , 2 \pi )$}), and the same formalism applies: The mean spectral density (in the variable $\theta \in [ 0 , 2 \pi )$) can be generically expanded in the Fourier series, \smash{$\rho_{\mathbf{U}} ( \theta ) = \frac{1}{2 \pi} ( 1 + \sum_{n \geq 1} ( m_{n} \ee^{- \ii n \theta} + m_{n}^{*} \ee^{\ii n \theta} ) )$}, where the coefficients \smash{$m_{n}$} are called ``moments.'' (The inverse Fourier transform and triangle inequality imply \smash{$| m_{n} | \leq 1$}.) Hence, the holomorphic Green function for $\mathbf{U}$,
\begin{equation}\label{eq:HolomorphicGreenFunctionForUnitary}
G_{\mathbf{U}} ( z ) = \int_{0}^{2 \pi} \frac{\rho_{\mathbf{U}} ( \theta ) \dd \theta}{z - \ee^{\ii \theta}} = \left\{ \begin{array}{ll} \frac{1 + M_{\mathbf{U}} ( z )}{z} , & \textrm{for } | z | > 1 , \\ - \frac{1}{z} M_{\mathbf{U}} \left( \frac{1}{z^{*}} \right)^{*} , & \textrm{for } | z | < 1 , \end{array} \right.
\end{equation}
where the positive moments are gathered into a generating function named the ``holomorphic $M$--transform,''
\begin{equation}\label{eq:HolomorphicMTransformForUnitary}
M_{\mathbf{U}} ( z ) \equiv \sum_{n \geq 1} \frac{m_{n}}{z^{n}} , \quad \textrm{for} \quad | z | > 1 .
\end{equation}


\subsubsection{Non--Hermitian Green function}
\label{sss:NonHermitianGreenFunction}

For non--Hermitian random matrices $\mathbf{X}$, the approach must be altogether different because the eigenvalues are generically complex, and in the large--$N$ limit occupy on average some two--dimensional domain $\mathcal{D}$. The mean spectral density is now defined through the complex Dirac delta (\ref{eq:NonHermitianMeanSpectralDensityDefinition}), whose representation \smash{$\delta^{( 2 )} ( \lambda - \lambda_{i} ) = \frac{1}{\pi} \partial_{\lambda^{*}} \lim_{\epsilon \to 0} \frac{\lambda^{*}}{| \lambda |^{2} + \epsilon^{2}}$} is at the roots of the definition of the ``non--holomorphic Green function''~\cite{SommersCrisantiSompolinskyStein1988,HaakeIzrailevLehmannSaherSommers1992,LehmannSaherSokolovSommers1995,FyodorovSommers1997,FyodorovKhoruzhenkoSommers1997},
\begin{equation}
\begin{split}\label{eq:NonHolomorphicGreenFunctionDefinition}
&G_{\mathbf{X}} ( z , z^{*} ) \equiv \lim_{\epsilon \to 0} \lim_{N \to \infty} \frac{1}{N} \sum_{i = 1}^{N} \la \frac{z^{*} - \lambda_{i}^{*}}{\left| z - \lambda_{i} \right|^{2} + \epsilon^{2}} \ra =\\
&= \lim_{\epsilon \to 0} \lim_{N \to \infty} \frac{1}{N} \Tr \la \frac{z^{*} \Id_{N} - \mathbf{X}^{\dagger}}{\left( z \Id_{N} - \mathbf{X} \right) \left( z^{*} \Id_{N} - \mathbf{X}^{\dagger} \right) + \epsilon^{2} \Id_{N}} \ra
\end{split}
\end{equation}
(with the convention for matrix division \smash{$\mathbf{A} / \mathbf{B} \equiv \mathbf{A} \mathbf{B}^{- 1}$}), since then the mean spectral density is obtained simply by taking a derivative,
\begin{equation}\label{eq:MeanSpectralDensityFromNonHolomorphicGreenFunction}
\rho_{\mathbf{X}} ( z , z^{*} ) = \frac{1}{\pi} \partial_{z^{*}} G_{\mathbf{X}} ( z , z^{*} ) , \quad \textrm{for} \quad z \in \mathcal{D} .
\end{equation}
(There are known intricacies concerning the order of limits in (\ref{eq:NonHolomorphicGreenFunctionDefinition}), but I will not be bothered by them.) An equivalent object, often handier, is the ``non--holomorphic $M$--transform,''
\begin{equation}\label{eq:NonHolomorphicMTransformDefinition}
M_{\mathbf{X}} ( z , z^{*} ) \equiv z G_{\mathbf{X}} ( z , z^{*} ) - 1 .
\end{equation}

\begin{figure}[t]
\includegraphics[width=\columnwidth]{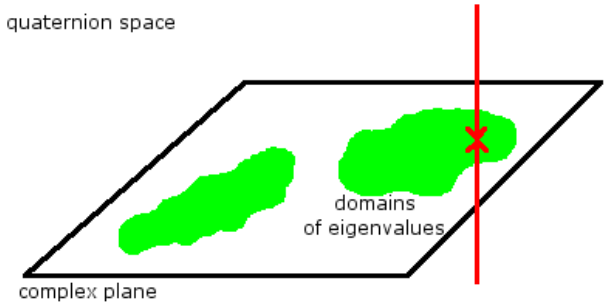}
\caption{For non--Hermitian random matrices, the eigenvalues are complex, and one needs a quaternion Green function of a quaternion variable.}
\label{fig:EigenvaluesComplex}
\end{figure}

The non--holomorphic Green function (\ref{eq:NonHolomorphicGreenFunctionDefinition}) is a more complicated object than the holomorphic counterpart (\ref{eq:HolomorphicGreenFunctionDefinition}) due to its denominator quadratic in $\mathbf{X}$. Hence, a linearizing procedure has been proposed~\cite{JanikNowakPappZahed1997-01} to introduce the ``matrix--valued Green function,'' which is a $2 \times 2$ matrix function of a complex variable,
\begin{equation}\label{eq:MatrixValuedGreenFunctionDefinition1}
\mathcal{G}_{\mathbf{X}} ( z , z^{*} ) \equiv \lim_{\epsilon \to 0} \lim_{N \to \infty} \frac{1}{N} \bTr \la \left( \mathcal{Z}_{\epsilon} \otimes \Id_{N} - \mathbf{X}^{\DD} \right)^{- 1} \ra ,
\end{equation}
where
\begin{equation}\label{eq:MatrixValuedGreenFunctionDefinition2}
\mathcal{Z}_{\epsilon} \equiv \left( \begin{array}{cc} z & \ii \epsilon \\ \ii \epsilon & z^{*} \end{array} \right) , \quad \mathbf{X}^{\DD} \equiv \left( \begin{array}{cc} \mathbf{X} & \Zero_{N} \\ \Zero_{N} & \mathbf{X}^{\dagger} \end{array} \right) ,
\end{equation}
while $\bTr$ (``block--trace'') turns a $2 N \times 2 N$ matrix into a $2 \times 2$ one by taking trace of its four $N \times N$ blocks,
\begin{equation}\label{eq:BlockTraceDefinition}
\bTr \left( \begin{array}{cc} \mathbf{A} & \mathbf{B} \\ \mathbf{C} & \mathbf{D} \end{array} \right) \equiv \left( \begin{array}{cc} \Tr \mathbf{A} & \Tr \mathbf{B} \\ \Tr \mathbf{C} & \Tr \mathbf{D} \end{array} \right) .
\end{equation}
This need to leave the complex plane --- into what we will see (paragraph~\ref{sss:NonHermitianQuaternionAdditionAlgorithm}) is the quaternion space --- is shown in figure~\ref{fig:EigenvaluesComplex}. Now, (\ref{eq:MatrixValuedGreenFunctionDefinition1}) is already linear in $\mathbf{X}$, with the structure mimicking that of the holomorphic Green function (\ref{eq:HolomorphicGreenFunctionDefinition}). Its upper left element is precisely the non--holomorphic Green function (\ref{eq:NonHolomorphicGreenFunctionDefinition}), \smash{$[ \mathcal{G}_{\mathbf{X}} ( z , z^{*} ) ]_{1 1} = G_{\mathbf{X}} ( z , z^{*} )$}. Its lower right element carries no new information, \smash{$[ \mathcal{G}_{\mathbf{X}} ( z , z^{*} ) ]_{2 2} = [ \mathcal{G}_{\mathbf{X}} ( z , z^{*} ) ]_{1 1}^{*}$}. Moreover, the negated product of the two off--diagonal elements (being a non--negative real number),
\begin{equation}\label{eq:CDefinition}
C_{\mathbf{X}} ( z , z^{*} ) \equiv - [ \mathcal{G}_{\mathbf{X}} ( z , z^{*} ) ]_{1 2} [ \mathcal{G}_{\mathbf{X}} ( z , z^{*} ) ]_{2 1} ,
\end{equation}
has been shown~\cite{ChalkerMehlig19982000,JanikNorenbergNowakPappZahed1999} to describe correlations between left and right eigenvectors of $\mathbf{X}$ (a property I will not exploit), and also plays a role of an ``order parameter'': it is positive inside the mean spectral domain $\mathcal{D}$ and zero outside of $\mathcal{D}$ (because \smash{$C_{\mathbf{X}} ( z , z^{*} ) \propto \epsilon^{2}$}, and the regulator $\epsilon$ can be set to zero outside of $\mathcal{D}$). From a practical point of view, once one finds \smash{$C_{\mathbf{X}} ( z , z^{*} )$} inside $\mathcal{D}$, then setting it to zero and solving for $z = x + \ii y$ yields an equation of the borderline $\partial \mathcal{D}$ in the Cartesian coordinates $( x , y )$.


\subsection{Quaternion free probability in a nutshell}
\label{ss:QuaternionFreeProbabilityInANutshell}

``Free probability,'' a theory initiated by Voiculescu and coworkers~\cite{VoiculescuDykemaNica1992} and Speicher~\cite{Speicher1994}, is a non--commutative probability theory (an instance of it being random matrix theory) endowed with a proper generalization of the classical notion of statistical independence, called ``freeness.'' Qualitatively, random matrices are free when not only are the entries of the distinct matrices statistically independent, but also when there is no angular correlation between them. I will not delve into details, just outline one important result of free probability, the ``addition algorithm.''


\subsubsection{Classical addition algorithm}
\label{sss:ClassicalAdditionAlgorithm}

In classical probability theory, if two random numbers \smash{$H_{1 , 2}$} are independent, then the PDF of their sum \smash{$( H_{1} + H_{2} )$} is derived using the ``classical addition algorithm'': First, the PDF's of the constituents are encoded into the ``characteristic functions,'' \smash{$g_{H_{1 , 2}} ( x ) \equiv \langle \ee^{\ii x H_{1 , 2}} \rangle$}, being complex functions of a real variable. Second, their logarithm is computed, \smash{$r_{H_{1 , 2}} ( x ) \equiv \log g_{H_{1 , 2}} ( x )$}. The independence property then ensures that this object simply adds when summing the random variables,
\begin{equation}\label{eq:ClassicalAdditionLaw}
r_{H_{1} + H_{2}} ( x ) = r_{H_{1}} ( x ) + r_{H_{2}} ( x ) .
\end{equation}
Third, exponentiating the result leads to the characteristic function, carrying the full spectral information, of the sum.


\subsubsection{Hermitian addition algorithm}
\label{sss:HermitianAdditionAlgorithm}

Free probability provides a similar algorithm for Hermitian random matrices: If \smash{$\mathbf{H}_{1 , 2}$} are free, then the mean spectral density of their sum \smash{$( \mathbf{H}_{1} + \mathbf{H}_{2} )$} is calculated as follows: First, one needs to have the holomorphic Green functions (\ref{eq:HolomorphicGreenFunctionDefinition}) of the constituents. Second, one computes their ``holomorphic Blue functions''~\cite{Zee1996}, being functional inverses of the holomorphic Green functions,
\begin{equation}\label{eq:HolomorphicBlueFunctionDefinition}
G_{\mathbf{H}_{1 , 2}} \left( B_{\mathbf{H}_{1 , 2}} ( z ) \right) = B_{\mathbf{H}_{1 , 2}} \left( G_{\mathbf{H}_{1 , 2}} ( z ) \right) = z .
\end{equation}
The freeness property then implies that these Blue functions obey
\begin{equation}\label{eq:HermitianAdditionLaw}
B_{\mathbf{H}_{1} + \mathbf{H}_{2}} ( z ) = B_{\mathbf{H}_{1}} ( z ) + B_{\mathbf{H}_{2}} ( z ) - \frac{1}{z} .
\end{equation}
Third, it remains to functionally invert the result to obtain the holomorphic Green function of the sum. (Let us mention that a more popular terminology is of the ``$R$--transform,'' \smash{$R_{\mathbf{H}} ( z ) \equiv B_{\mathbf{H}} ( z ) - 1 / z$}, which is simply additive, \smash{$R_{\mathbf{H}_{1} + \mathbf{H}_{2}} ( z ) = R_{\mathbf{H}_{1}} ( z ) + R_{\mathbf{H}_{2}} ( z )$}.)


\subsubsection{Non--Hermitian (quaternion) addition algorithm}
\label{sss:NonHermitianQuaternionAdditionAlgorithm}

In~\cite{JaroszNowak2004,JaroszNowak2006}, a straightforward extension to the non--Hermitian world has been proposed. The procedure lies on the observation that the matrix--valued Green function (\ref{eq:MatrixValuedGreenFunctionDefinition1}) looks analogously to the holomorphic Green function (\ref{eq:HolomorphicGreenFunctionDefinition}) considered in the vicinity of the mean eigenvalue cuts, in the imaginary direction, \ie \smash{$G_{\mathbf{H}} ( \lambda + \ii \epsilon )$} --- while in order to compute the holomorphic Blue function (\ref{eq:HolomorphicBlueFunctionDefinition}), the holomorphic Green function must be known on the whole complex plane. Therefore, the authors of~\cite{JaroszNowak2004,JaroszNowak2006} felt compelled to replace the infinitesimally small regulator $\epsilon$ in (\ref{eq:MatrixValuedGreenFunctionDefinition1}) by an arbitrary complex number $d$ (actually, real and non--negative $d$ would be sufficient, since $\epsilon$ is real and non--negative),
\begin{equation}\label{eq:QuaternionDefinition}
\mathcal{Z}_{\epsilon} = \left( \begin{array}{cc} z & \ii \epsilon \\ \ii \epsilon & z^{*} \end{array} \right) \quad \to \quad \mathcal{Q} \equiv \left( \begin{array}{cc} c & \ii d^{*} \\ \ii d & c^{*} \end{array} \right)
\end{equation}
(where $c$ is also an arbitrary complex number), thereby defining the ``quaternion Green function'' as a quaternion function of a quaternion variable,
\begin{equation}\label{eq:QuaternionGreenFunctionDefinition}
\mathcal{G}_{\mathbf{X}} ( \mathcal{Q} ) \equiv \left( \begin{array}{cc} a & \ii b^{*} \\ \ii b & a^{*} \end{array} \right) \equiv \frac{1}{N} \bTr \la \left( \mathcal{Q} \otimes \Id_{N} - \mathbf{X}^{\DD} \right)^{- 1} \ra .
\end{equation}
(I will henceforth refer to $c$, $d$ and $a$, $b$ as to the ``coefficients'' of the respective quaternion.)

The ``quaternion addition algorithm'' has then been proven: Let \smash{$\mathbf{X}_{1 , 2}$} be free non--Hermitian random matrices. First, their quaternion Green functions (\ref{eq:QuaternionGreenFunctionDefinition}) must be found (see subsection~\ref{ss:RationalHermitizationProcedure} for how it is done in the case of Hermitian or unitary matrices). Second, these quaternion Green functions are to be functionally inverted in the quaternion space, leading to the ``quaternion Blue functions,''
\begin{equation}\label{eq:QuaternionBlueFunctionDefinition}
\mathcal{G}_{\mathbf{X}_{1 , 2}} \left( \mathcal{B}_{\mathbf{X}_{1 , 2}} ( \mathcal{Q} ) \right) = \mathcal{B}_{\mathbf{X}_{1 , 2}} \left( \mathcal{G}_{\mathbf{X}_{1 , 2}} ( \mathcal{Q} ) \right) = \mathcal{Q} .
\end{equation}
The freeness of the summands suffices to show that an analogue of (\ref{eq:HermitianAdditionLaw}) holds at the quaternion level,
\begin{equation}\label{eq:QuaternionAdditionLaw}
\mathcal{B}_{\mathbf{X}_{1} + \mathbf{X}_{2}} ( \mathcal{Q} ) = \mathcal{B}_{\mathbf{X}_{1}} ( \mathcal{Q} ) + \mathcal{B}_{\mathbf{X}_{2}} ( \mathcal{Q} ) - \mathcal{Q}^{- 1} .
\end{equation}
Third, it is enough to functionally invert the result at the point \smash{$\mathcal{Z}_{\epsilon = 0}$} to obtain the matrix--valued Green function (\ref{eq:MatrixValuedGreenFunctionDefinition1}) of the sum,
\begin{equation}\label{eq:MatrixValuedGreenFunctionFromQuaternionBlueFunction}
\mathcal{B}_{\mathbf{X}_{1} + \mathbf{X}_{2}} \left( \left( \begin{array}{cc} a & \ii b^{*} \\ \ii b & a^{*} \end{array} \right) \right) = \left( \begin{array}{cc} z & 0 \\ 0 & z^{*} \end{array} \right) ,
\end{equation}
where \smash{$a = G_{\mathbf{X}_{1} + \mathbf{X}_{2}} ( z , z^{*} )$} and \smash{$| b |^{2} = C_{\mathbf{X}_{1} + \mathbf{X}_{2}} ( z , z^{*} )$}. (The regulator $\epsilon$ may be set to zero here because it will be seen that the functional inversion to be performed in equation (\ref{eq:MatrixValuedGreenFunctionFromQuaternionBlueFunction}) takes care by itself of regulating the singularities at the eigenvalues: (\ref{eq:MatrixValuedGreenFunctionFromQuaternionBlueFunction}) will always yield a ``holomorphic solution,'' valid outside of $\mathcal{D}$, and a ``non--holomorphic solution,'' inside $\mathcal{D}$.)


\subsection{``Rational Hermitization'' procedure}
\label{ss:RationalHermitizationProcedure}


\subsubsection{Description of the method}
\label{sss:DescriptionOfTheMethod}

There exists a special class of random matrices which permit an explicit calculation of the quaternion Green function (\ref{eq:QuaternionGreenFunctionDefinition}) by means of a procedure called ``rational Hermitization.'' To describe it, perform the matrix inversion in (\ref{eq:QuaternionGreenFunctionDefinition}), which gives the coefficients $a$, $b$ of the quaternion Green function through the coefficients $c$, $d$ of its argument,
\begin{subequations}
\begin{align}
a =& \frac{1}{N} \Tr \la \frac{c^{*} \Id_{N} - \mathbf{X}^{\dagger}}{\mathbf{X} \mathbf{X}^{\dagger} - c \mathbf{X}^{\dagger} - c^{*} \mathbf{X} + \left( | c |^{2} + | d |^{2} \right) \Id_{N}} \ra ,\label{eq:QuaternionGreenFunctiona}\\
b =& \frac{1}{N} \Tr \la \frac{- d}{\mathbf{X} \mathbf{X}^{\dagger} - c \mathbf{X}^{\dagger} - c^{*} \mathbf{X} + \left( | c |^{2} + | d |^{2} \right) \Id_{N}} \ra .\label{eq:QuaternionGreenFunctionb}
\end{align}
\end{subequations}
Consider a symmetry constraint on $\mathbf{X}$ such that \smash{$\mathbf{X}^{\dagger}$} is a rational function of $\mathbf{X}$. Two primary examples are Hermitian (\smash{$\mathbf{H}^{\dagger} = \mathbf{H}$}) and unitary (\smash{$\mathbf{U}^{\dagger} = \mathbf{U}^{- 1}$}) matrices. Then on the RHS of (\ref{eq:QuaternionGreenFunctiona})--(\ref{eq:QuaternionGreenFunctionb}) one obtains rational functions of $\mathbf{X}$, which in turn may be expanded into simple fractions, and therefore written in terms of the holomorphic Green function (\ref{eq:HolomorphicGreenFunctionDefinition}) of $\mathbf{X}$. In other words, for this class of random matrices, knowing the holomorphic Green function is sufficient for knowing the (more complicated) quaternion Green function.

The rational Hermitization procedure for Hermitian matrices has been outlined in~\cite{JaroszNowak2004,JaroszNowak2006} (it is needed \eg when one adds a Hermitian to a non--Hermitian random matrix).


\subsubsection{Application to unitary random matrices}
\label{sss:ApplicationToUnitaryRandomMatrices}

For an arbitrary unitary random matrix $w \mathbf{U}$ (where for further convenience I have included an arbitrary complex number $w$), one gets in this way the following coefficients of the quaternion Green function,
\begin{subequations}
\begin{align}
a =& \frac{1}{2 c} \left( \frac{- | w |^{2} + | c |^{2} - | d |^{2}}{g} + 1 \right) \left( 1 + M_{\mathbf{U}} ( u ) \right) +\nonumber\\
&+ \frac{1}{2 c} \left( \frac{- | w |^{2} + | c |^{2} - | d |^{2}}{g} - 1 \right) M_{\mathbf{U}} ( u )^{*} ,\label{eq:QuaternionGreenFunctionaForUnitary}\\
b =& - \frac{d}{g} \left( 1 + M_{\mathbf{U}} ( u ) + M_{\mathbf{U}} ( u )^{*} \right) ,\label{eq:QuaternionGreenFunctionbForUnitary}
\end{align}
\end{subequations}
where \smash{$g \equiv \sqrt{( ( | c | - | w | )^{2} + | d |^{2} ) ( ( | c | + | w | )^{2} + | d |^{2} )}$} and \smash{$u \equiv \frac{1}{2 w \overline{c}} ( | w |^{2} + | c |^{2} + | d |^{2} + g )$}, and recall the holomorphic $M$--transform of $\mathbf{U}$ (\ref{eq:HolomorphicMTransformForUnitary}).


\section{Summing free unitary random matrices --- the eigenvalues}
\label{s:EIGSummingFreeUnitaryRandomMatrices}

After the self--contained introduction in sections~\ref{s:Introduction} and~\ref{s:QuaternionFreeProbability}, I now come to the first main objective of this work, \ie considering the model $\mathbf{S}$ (\ref{eq:SDefinition}) and exploiting the quaternion addition law (\ref{eq:QuaternionAdditionLaw}),
\begin{equation}\label{eq:QuaternionAdditionLawForS}
\mathcal{B}_{\mathbf{S}} ( \mathcal{Q} ) = \mathcal{B}_{w_{1} \mathbf{U}_{1}} ( \mathcal{Q} ) + \ldots + \mathcal{B}_{w_{L} \mathbf{U}_{L}} ( \mathcal{Q} ) - ( L - 1 ) \mathcal{Q}^{- 1} ,
\end{equation}
to calculate (\ref{eq:MatrixValuedGreenFunctionFromQuaternionBlueFunction}) the non--holomorphic Green function (\ref{eq:NonHolomorphicGreenFunctionDefinition}) of $\mathbf{S}$, and consequently (\ref{eq:MeanSpectralDensityFromNonHolomorphicGreenFunction}) its mean spectral density.


\subsection{Summing free CUE random matrices}
\label{ss:EIGSummingFreeCUERandomMatrices}

Let us commence from solving this problem for all the \smash{$\mathbf{U}_{l}$}'s belonging to the simplest unitary ensemble, the ``circular unitary ensemble'' (CUE), defined to have all the moments zero, \smash{$m_{n} = 0$}, \ie \smash{$M_{\mathbf{U}} ( u ) = 0$}, \ie the mean spectral density constant on $C ( 0 , 1 )$, \smash{$\rho_{\mathbf{CUE}} ( \theta ) = 1 / 2 \pi$}.


\subsubsection{The master equation}
\label{sss:EIGTheMasterEquation}

In this case, equations (\ref{eq:QuaternionGreenFunctionaForUnitary})--(\ref{eq:QuaternionGreenFunctionbForUnitary}) can be solved to provide the coefficients $c$, $d$ of the quaternion Blue function \smash{$\mathcal{B}_{w \mathbf{CUE}} ( \mathcal{Q} )$} through the coefficients $a$, $b$ of the quaternion $\mathcal{Q}$,
\begin{subequations}
\begin{align}
c &= \frac{a^{*}}{| a |^{2} + | b |^{2}} ,\label{eq:QuaternionBlueFunctioncForCUE}\\
d &= - b \left( \frac{1}{| a |^{2} + | b |^{2}} - \frac{1 + s \sqrt{1 - 4 | w |^{2} | b |^{2}}}{2 | b |^{2}} \right) ,\label{eq:QuaternionBlueFunctiondForCUE}
\end{align}
\end{subequations}
for some sign $s = \pm 1$.

For the sum (\ref{eq:SDefinition}) of the CUE's with general weights \smash{$w_{l}$}, the quaternion addition law (\ref{eq:QuaternionAdditionLawForS}) along with (\ref{eq:MatrixValuedGreenFunctionFromQuaternionBlueFunction}) yield equations for the coefficients $a$, $b$ of the matrix--valued Green function (\ref{eq:MatrixValuedGreenFunctionDefinition1}) of $\mathbf{S}$, \ie \smash{$a = G_{\mathbf{S}} ( z , z^{*} )$} (or better (\ref{eq:NonHolomorphicMTransformDefinition}), \smash{$M \equiv M_{\mathbf{S}} ( z , z^{*} ) = z a - 1$}) and \smash{$| b |^{2} = C_{\mathbf{S}} ( z , z^{*} )$},
\begin{subequations}
\begin{align}
z &= c_{1} + \ldots + c_{L} - \frac{( L - 1 ) a^{*}}{| a |^{2} + | b |^{2}} ,\label{eq:QuaternionAdditionLawcForSWithCUEs}\\
0 &= d_{1} + \ldots + d_{L} + \frac{( L - 1 ) b}{| a |^{2} + | b |^{2}} ,\label{eq:QuaternionAdditionLawdForSWithCUEs}
\end{align}
\end{subequations}
where the \smash{$c_{l}$}'s and \smash{$d_{l}$}'s are given by (\ref{eq:QuaternionBlueFunctioncForCUE})--(\ref{eq:QuaternionBlueFunctiondForCUE}).

This set (\ref{eq:QuaternionBlueFunctioncForCUE}), (\ref{eq:QuaternionBlueFunctiondForCUE}), (\ref{eq:QuaternionAdditionLawcForSWithCUEs}), (\ref{eq:QuaternionAdditionLawdForSWithCUEs}) can be easily simplified to the following form: $M$ is a solution to the equation
\begin{equation}\label{eq:MainEquationForMForSWithCUEs}
L + 2 M = \sum_{l = 1}^{L} s_{l} \sqrt{1 + 4 \left| w_{l} \right|^{2} \frac{M ( M + 1 )}{| z |^{2}}} ,
\end{equation}
for some proper choice of the signs \smash{$s_{l}$} (yielding a meaningful solution), and then also
\begin{equation}\label{eq:CThroughMForSWithCUEs}
C_{\mathbf{S}} ( z , z^{*} ) = - \frac{M ( M + 1 )}{| z |^{2}} .
\end{equation}
This is the master equation for summing free CUE random matrices.

\begin{description}
\item[Remark 1:] The master equation (\ref{eq:MainEquationForMForSWithCUEs}) and formula (\ref{eq:CThroughMForSWithCUEs}) may be cast in a more elegant way,
    \begin{equation}\label{eq:MainEquationForMForSWithCUEsVersion2Equation1}
    M = \sum_{l = 1}^{L} M_{l} ,
    \end{equation}
    where
    \begin{subequations}
    \begin{align}
    - \frac{M ( M + 1 )}{| z |^{2}} &= C ,\label{eq:MainEquationForMForSWithCUEsVersion2Equation2}\\
    - \frac{M_{l} ( M_{l} + 1 )}{\left| w_{l} \right|^{2}} &= C , \quad l = 1 , 2 , \ldots , L ,\label{eq:MainEquationForMForSWithCUEsVersion2Equation3}
    \end{align}
    \end{subequations}
    where for short \smash{$C \equiv C_{\mathbf{S}} ( z , z^{*} )$}. This form is surprisingly symmetric: Equation (\ref{eq:MainEquationForMForSWithCUEsVersion2Equation2}) is identical as the set of definitions (\ref{eq:MainEquationForMForSWithCUEsVersion2Equation3}) (\emph{cf.} (\ref{eq:SVMainEquationForGForSWithCUEsVersion2Equation1})--(\ref{eq:SVMainEquationForGForSWithCUEsVersion2Equation3})).
\item[Remark 2:] The above set (\ref{eq:MainEquationForMForSWithCUEsVersion2Equation1})--(\ref{eq:MainEquationForMForSWithCUEsVersion2Equation3}) also provides an algorithm for rewriting it as a single polynomial equation for $M$: (1) Find \smash{$M_{L}$} from (\ref{eq:MainEquationForMForSWithCUEsVersion2Equation1}). (2) Substitute it to (\ref{eq:MainEquationForMForSWithCUEsVersion2Equation3}), $l = L$. (3) Remove from it all the squares \smash{$M_{l}^{2}$}, $l = 1 , \ldots , L - 1$, by using (\ref{eq:MainEquationForMForSWithCUEsVersion2Equation3}), $l = 1 , \ldots , L - 1$. (4) Find from it \smash{$M_{L - 1}$} (this requires solving a linear equation), and return to step (2), with $l = L - 1$. The procedure ends at \smash{$M_{0} \equiv M$}. One may also note that all the coefficients of the resulting polynomial equation are themselves symmetric polynomials in the $( L + 1 )$ arguments \smash{$| w_{l} |^{2}$}, $l = 0 , 1 , \ldots , L$, where \smash{$w_{0} \equiv z$}. However, in all the five examples presented below, it is better not to use this algorithm, but rather successively square the square roots in (\ref{eq:MainEquationForMForSWithCUEs}), as this leads to a polynomial equation of a lower order. For more general cases, however, the former algorithm seems inevitable.
\item[Remark 3:] $M$ depends only on
    \begin{equation}\label{eq:RDefinition}
    R \equiv | z | ,
    \end{equation}
    \ie the mean spectral density (\ref{eq:MeanSpectralDensityFromNonHolomorphicGreenFunction}) is rotationally--symmetric around zero, and may be expressed as \smash{$\rho_{\mathbf{S}} ( z , z^{*} ) = \frac{1}{\pi} \partial_{R^{2}} M$}. Hence, we may focus on investigating just its radial part,
    \begin{equation}\label{eq:RadialMeanSpectralDensityFromRotationallySymmetricNonHolomorphicMTransform}
    \rho^{\textrm{rad}}_{\mathbf{S}} ( R ) \equiv 2 \pi R \left. \rho_{\mathbf{S}} ( z , z^{*} ) \right|_{| z | = R} = \frac{\dd M}{\dd R} .
    \end{equation}
\item[Remark 4:] Only the absolute values, and not the complex arguments, of the weights \smash{$w_{l}$} are relevant. Hence, it is enough to study real and positive weights.
\item[Remark 5:] As explained at the end of paragraph~\ref{sss:NonHermitianGreenFunction}, once $C$ is found, then $C = 0$ is the equation of the borderline $\partial \mathcal{D}$ of the mean spectral domain. According to (\ref{eq:CThroughMForSWithCUEs}) (or (\ref{eq:MainEquationForMForSWithCUEsVersion2Equation2})), this happens when $M = 0$ or $M = - 1$, \ie in terms of the Green function (\ref{eq:NonHolomorphicMTransformDefinition}), $G = 1 / z$ or $G = 0$. This represents the matching on $\partial \mathcal{D}$ of the non--holomorphic quantities, valid inside $\mathcal{D}$ (\ie $C$, $M$, $G$), with the holomorphic quantities, valid outside $\mathcal{D}$ (\ie respectively, $0$, $0$ or $- 1$, $1 / z$ or $0$). This means that the mean spectral domain $\mathcal{D}$ is either (1) a centered disk, enclosed by the circle \smash{$R = R_{\textrm{ext}}$} --- where the radius \smash{$R_{\textrm{ext}}$} is found by substituting $M = 0$ to the master equation --- or (2) a centered annulus, confined by an external circle \smash{$R = R_{\textrm{ext}}$} ($M = 0$) and an internal one \smash{$R = R_{\textrm{int}}$} ($M = - 1$). This is an instance of the ``Feinberg--Zee single ring theorem''~\cite{FeinbergZee1997-02,FeinbergScalettarZee2001,Feinberg2006,GuionnetKrishnapurZeitouni2009}, which states that if the mean spectral density is rotationally--symmetric around zero, then $\mathcal{D}$ is either a disk or an annulus.
\end{description}

I will now explicitly solve the master equation (\ref{eq:MainEquationForMForSWithCUEs}), or reduce it to a polynomial equation, in five cases.


\subsubsection{Example 1: $L = 2$}
\label{sss:EIGExample1}

\begin{figure}[t]
\includegraphics[width=\columnwidth]{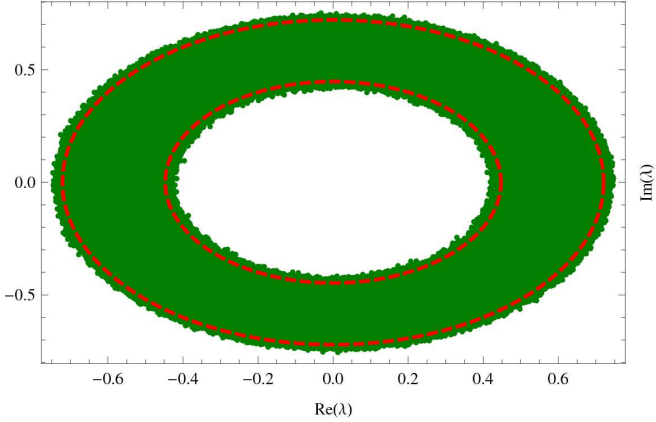}
\caption{The \smash{$5 \cdot 10^{5}$} eigenvalues of the sum of $L = 2$ CUE random matrices of dimension $N = 500$, weighted with \smash{$w_{1} = 0.4$} and \smash{$w_{2} = 1 - w_{1} = 0.6$}, obtained from $1,\!000$ Monte--Carlo iterations (the points), and the theoretical annulus (\ref{eq:RExtForSWithTwoCUEs})--(\ref{eq:RIntForSWithTwoCUEs}) (the dashed lines).}
\label{fig:EIGSumTWOUnitaryIter1000N500ReIm}
\end{figure}

\begin{figure*}[t]
\includegraphics[width=\columnwidth]{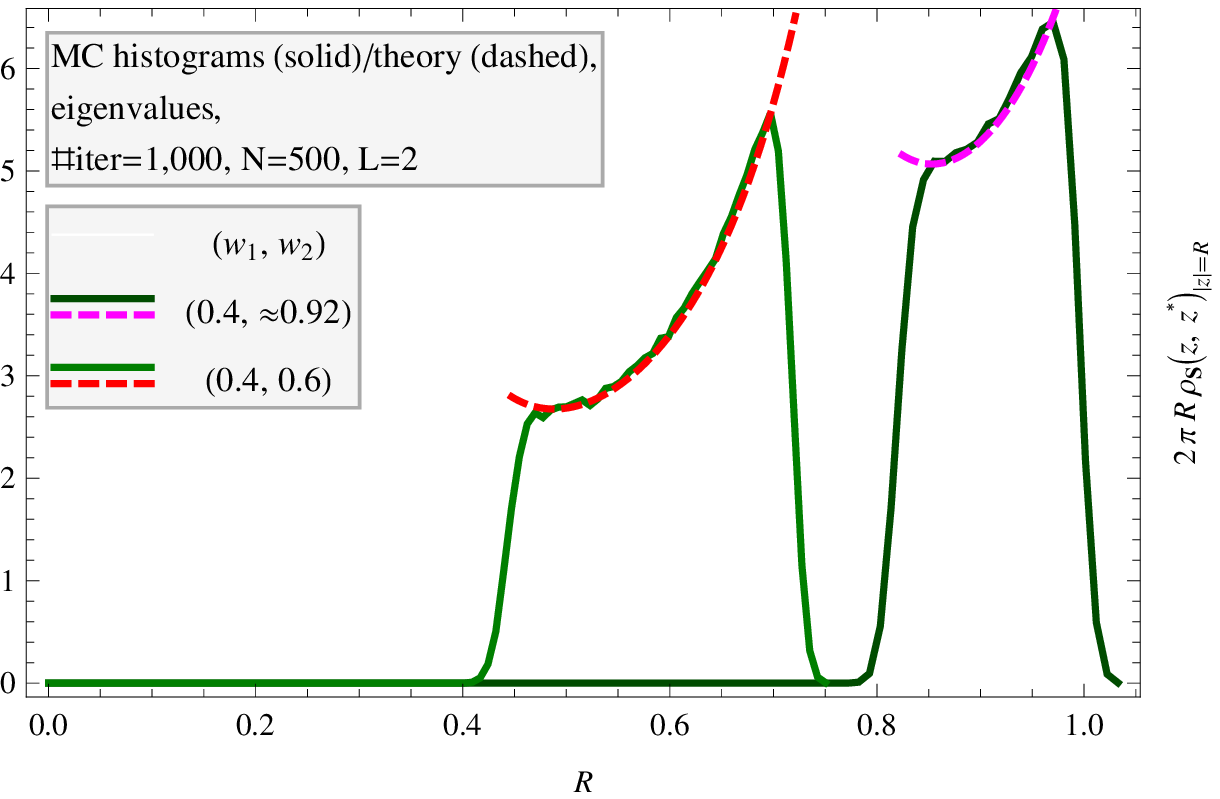}
\includegraphics[width=\columnwidth]{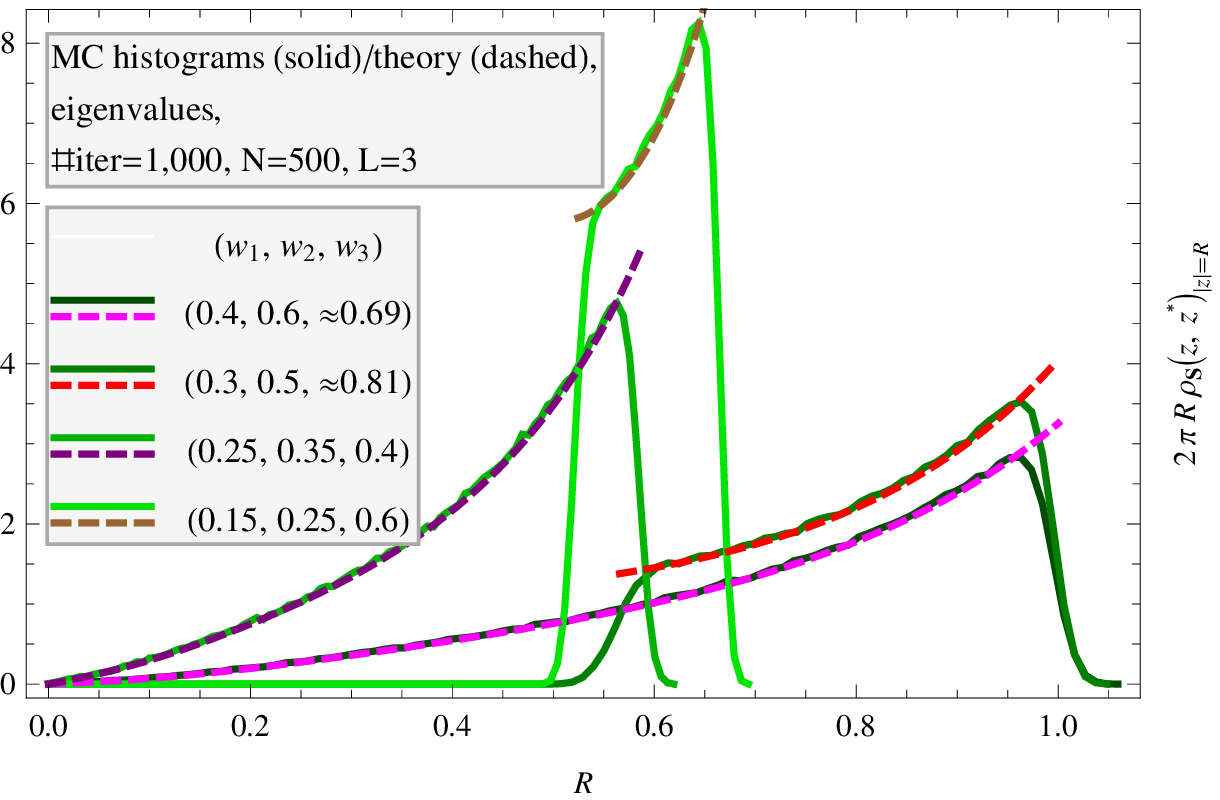}
\includegraphics[width=\columnwidth]{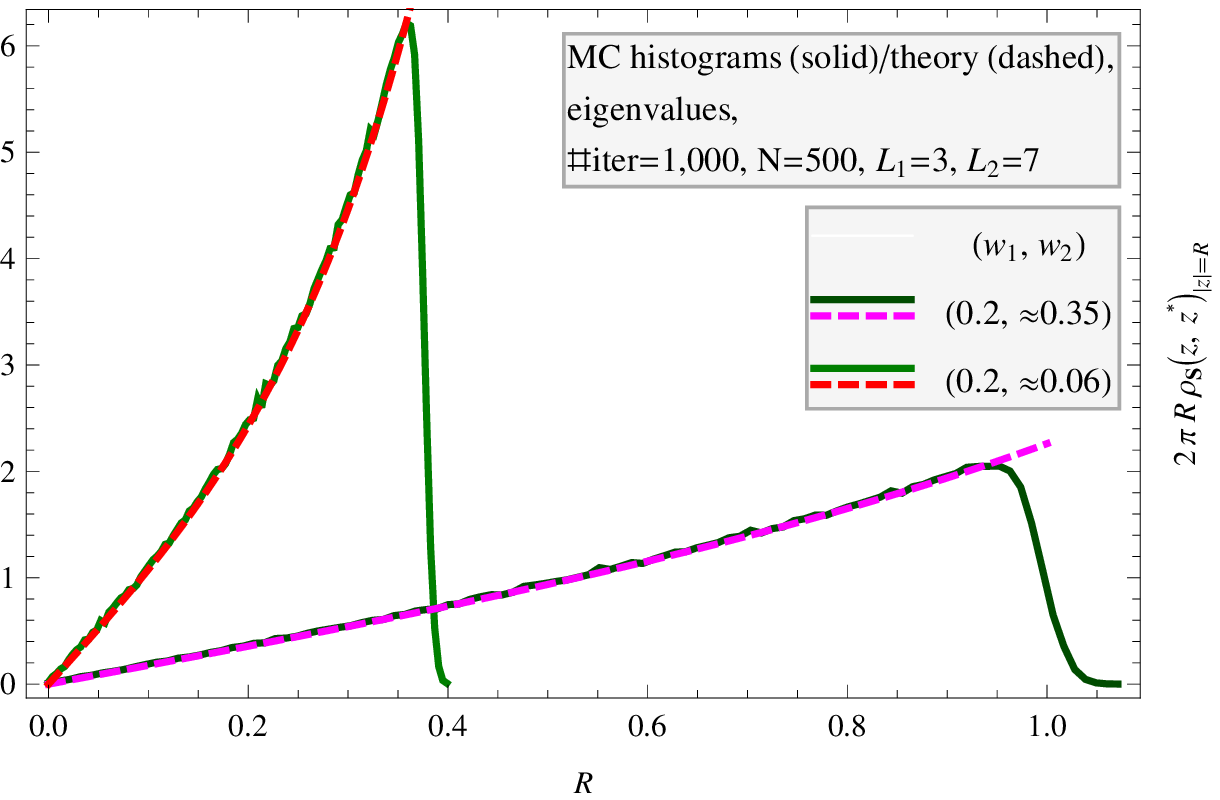}
\includegraphics[width=\columnwidth]{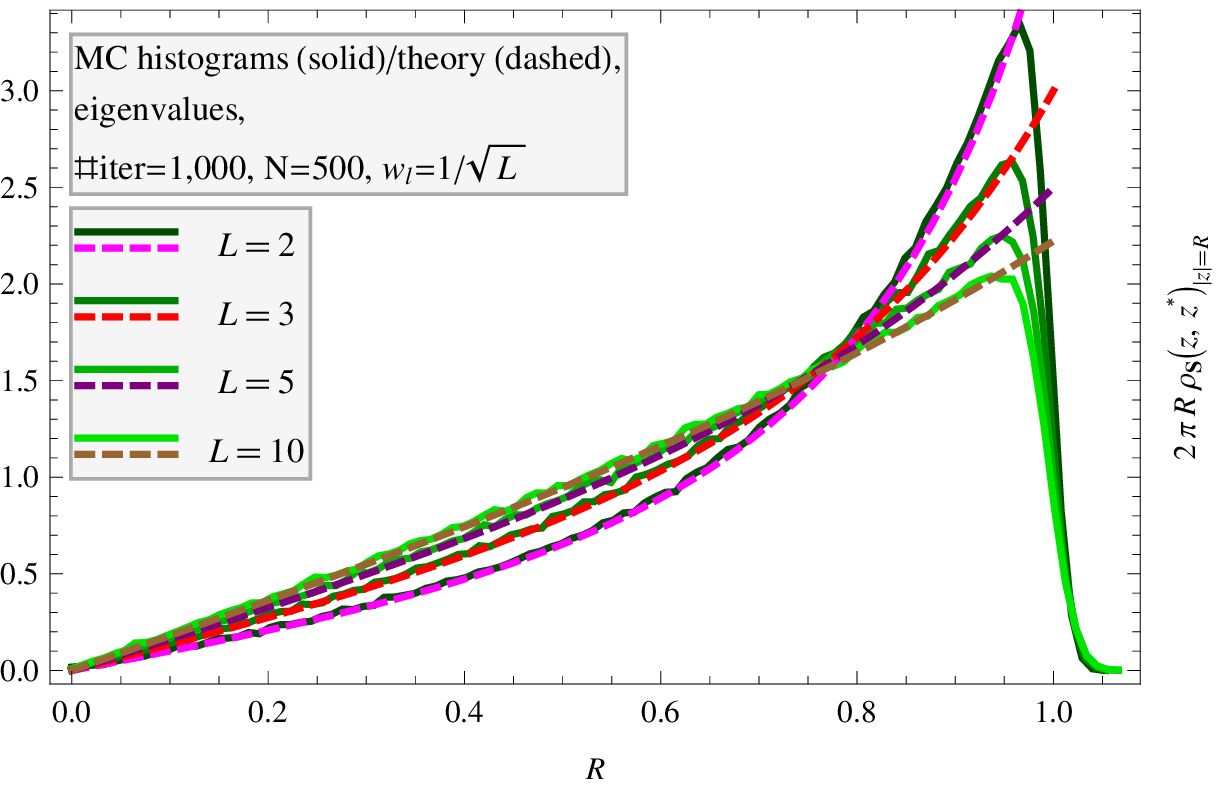}
\caption{The radial part of the mean spectral density, \smash{$\rho^{\textrm{rad}}_{\mathbf{S}} ( R )$} (\ref{eq:RadialMeanSpectralDensityFromRotationallySymmetricNonHolomorphicMTransform}), for (left to right, top to bottom):\\
A: The sum of $L = 2$ CUE random matrices weighted with \smash{$w_{1} = 0.4$} and (1) \smash{$w_{2} = \sqrt{1 - w_{1}^{2}} \approx 0.92$}, (2) \smash{$w_{2} = 1 - w_{1} = 0.6$}.\\
B: The sum of $L = 3$ CUE random matrices weighted with (1) \smash{$w_{1} = 0.4$}, \smash{$w_{2} = 0.6$}, \smash{$w_{3} = \sqrt{1 - w_{1}^{2} - w_{2}^{2}} \approx 0.69$} ($\mathcal{D}$ is a disk), (2) \smash{$w_{1} = 0.3$}, \smash{$w_{2} = 0.5$}, \smash{$w_{3} = \sqrt{1 - w_{1}^{2} - w_{2}^{2}} \approx 0.81$} ($\mathcal{D}$ is an annulus), (3) \smash{$w_{1} = 0.25$}, \smash{$w_{2} = 0.35$}, \smash{$w_{3} = 1 - w_{1} - w_{2} = 0.4$} ($\mathcal{D}$ is a disk), (4) \smash{$w_{1} = 0.15$}, \smash{$w_{2} = 0.25$}, \smash{$w_{3} = 1 - w_{1} - w_{2} = 0.6$} ($\mathcal{D}$ is an annulus).\\
C: The sum of $L = 10$ CUE random matrices such that \smash{$L_{1} = 3$} weights equal \smash{$w_{1}$}, and \smash{$L_{2} = 7$} weights equal \smash{$w_{2}$}, where \smash{$w_{1} = 0.2$} and (1) \smash{$w_{2} = \sqrt{( 1 - L_{1} w_{1}^{2} ) / L_{2}} \approx 0.35$}, (2) \smash{$w_{2} = ( 1 - L_{1} w_{1} ) / L_{2} \approx 0.06$}.\\
D: The sum of $L = 2$, $3$, $5$, $10$ CUE random matrices weighted with the same value \smash{$w = 1 / \sqrt{L}$}. One observes that as $L$ grows, the bulk plots approach the straight line $R \to 2 R$, corresponding to the GinUE.\\
The theoretical results (\ref{eq:RhoForSWithTwoCUEs}), (\ref{eq:MForSWithThreeCUEs}), (\ref{eq:MForSWithL1L2CUEs}), (\ref{eq:RhoForSWithLCUEs}), respectively (dashed lines) checked against numerical Monte--Carlo simulations, with matrix dimension $N = 500$, $1,\!000$ iterations, and $100$--bin histograms (solid lines).}
\label{fig:EIG}
\end{figure*}

As a first example of using the master equation (\ref{eq:MainEquationForMForSWithCUEs}), consider a sum of $L = 2$ free CUE matrices, with arbitrary weights \smash{$w_{1 , 2}$}. The master equation becomes linear, and yields the non--holomorphic solution,
\begin{equation}\label{eq:MForSWithTwoCUEs}
M = \frac{1}{\frac{\left( \left| w_{1} \right| + \left| w_{2} \right| \right)^{2}}{R^{2}} - 1} + \frac{1}{\frac{\left( \left| w_{1} \right| - \left| w_{2} \right| \right)^{2}}{R^{2}} - 1} .
\end{equation}

Both holomorphic solutions, $M = 0$ and $M = - 1$, are compatible with (\ref{eq:MForSWithTwoCUEs}), which implies that $\mathcal{D}$ is a centered annulus of radii,
\begin{subequations}
\begin{align}
R_{\textrm{ext}} &= \sqrt{| w_{1} |^{2} + | w_{2} |^{2}} ,\label{eq:RExtForSWithTwoCUEs}\\
R_{\textrm{int}} &= \sqrt{| | w_{1} |^{2} - | w_{2} |^{2} |} .\label{eq:RIntForSWithTwoCUEs}
\end{align}
\end{subequations}
(It becomes a disk only when \smash{$w_{1} = w_{2}$}, see figure~\ref{fig:EIG} (D).)

The radial mean spectral density of $\mathbf{S}$ follows by applying (\ref{eq:RadialMeanSpectralDensityFromRotationallySymmetricNonHolomorphicMTransform}) to (\ref{eq:MForSWithTwoCUEs}),
\begin{equation}
\begin{split}\label{eq:RhoForSWithTwoCUEs}
&\rho^{\textrm{rad}}_{\mathbf{S}} ( R ) = 2 R \cdot\\
&\cdot \left( \frac{\frac{1}{\left( \left| w_{1} \right| + \left| w_{2} \right| \right)^{2}}}{\left( \frac{R^{2}}{\left( \left| w_{1} \right| + \left| w_{2} \right| \right)^{2}} - 1 \right)^{2}} + \frac{\frac{1}{\left( \left| w_{1} \right| - \left| w_{2} \right| \right)^{2}}}{\left( \frac{R^{2}}{\left( \left| w_{1} \right| - \left| w_{2} \right| \right)^{2}} - 1 \right)^{2}} \right) ,
\end{split}
\end{equation}
for \smash{$R_{\textrm{int}} \leq R \leq R_{\textrm{ext}}$}, and zero otherwise. (The apparent singularities lie outside of the annulus.) I show the experimental eigenvalues on the complex plane and mark the theoretical annulus in figure~\ref{fig:EIGSumTWOUnitaryIter1000N500ReIm}. Formula (\ref{eq:RhoForSWithTwoCUEs}) is numerically checked in figure~\ref{fig:EIG} (A), finding perfect agreement in the bulk.


\subsubsection{Example 2: $L = 3$}
\label{sss:EIGExample2}

Consider now a sum of $L = 3$ free CUE matrices, with arbitrary weights \smash{$w_{1 , 2 , 3}$}. The master equation can be transformed into a fourth--order polynomial equation,
\begin{equation}
\begin{split}\label{eq:MForSWithThreeCUEs}
&M^{4} \left( R^{2} - v_{0}^{2} \right) \left( R^{2} - v_{1}^{2} \right) \left( R^{2} - v_{2}^{2} \right) \left( R^{2} - v_{3}^{2} \right) +\\
&+ M^{3} \Big( 9 R^{8} - 28 \mu_{( 1 )} R^{6} + 10 \left( 3 \mu_{( 2 )} + 2 \mu_{( 1 , 1 )} \right) R^{4} +\\
&+ 12 \left( - \mu_{( 3 )} + \mu_{( 2 , 1 )} - 10 \mu_{( 1 , 1 , 1 )} \right) R^{2} + v_{0}^{2} v_{1}^{2} v_{2}^{2} v_{3}^{2} \Big) +\\
&+ 2 M^{2} R^{2} \Big( 15 R^{6} - 35 \mu_{( 1 )} R^{4} + \left( 25 \mu_{( 2 )} + 14 \mu_{( 1 , 1 )} \right) R^{2} -\\
&- 5 \mu_{( 3 )} + 5 \mu_{( 2 , 1 )} - 42 \mu_{( 1 , 1 , 1 )} \Big) +\\
&+ 4 M R^{4} \left( 11 R^{4} - 18 \mu_{( 1 )} R^{2} + 7 \mu_{( 2 )} + 2 \mu_{( 1 , 1 )} \right) +\\
&+ 24 R^{6} \left( R^{2} - \mu_{( 1 )} \right) = 0 ,
\end{split}
\end{equation}
where the basis of the monomial symmetric polynomials,
\begin{equation}\label{eq:MonomialSymmetricPolynomials}
\mu_{p} \equiv \sum_{\pi \in \textrm{Perm} ( p )} \sum_{\gamma \in \textrm{Subs} ( L , \# p )} \prod_{s = 1}^{\# p} | w_{\gamma ( s )} |^{2 \pi ( s )} ,
\end{equation}
where $p$ is a partition, $\textrm{Perm} ( p )$ the set of all its permutations, $\textrm{Subs} ( L , \# p )$ the set of all the $\# p$--element subsets of $\{ 1 , \ldots , L \}$; also \smash{$v_{l} \equiv \sum_{k = 1}^{L} ( 1 - 2 \delta_{k l} ) | w_{k} |$}.

Inserting $M = 0$ or $M = - 1$ into (\ref{eq:MForSWithThreeCUEs}) leads to the values of the radii of the circles enclosing the mean spectral domain $\mathcal{D}$,
\begin{subequations}
\begin{align}
R_{\textrm{ext}} &= \sqrt{| w_{1} |^{2} + | w_{2} |^{2} + | w_{3} |^{2}} ,\label{eq:RExtForSWithThreeCUEs}\\
R_{\textrm{int}} &= \sqrt{\max \left( - V_{1} , - V_{2} , - V_{3} , 0 \right)} ,\label{eq:RIntForSWithThreeCUEs}
\end{align}
\end{subequations}
where \smash{$V_{l} \equiv \sum_{k = 1}^{L} ( 1 - 2 \delta_{k l} ) | w_{k} |^{2}$}. One may check that the three numbers, \smash{$- V_{1 , 2 , 3}$}, are either all negative or only one is positive; accordingly, $\mathcal{D}$ is a disk or an annulus.

I put forward a hypothesis that formulae analogous to (\ref{eq:RExtForSWithThreeCUEs})--(\ref{eq:RIntForSWithThreeCUEs}) hold for arbitrary $L$.

The radial mean spectral density is given by (\ref{eq:RadialMeanSpectralDensityFromRotationallySymmetricNonHolomorphicMTransform}) and equation (\ref{eq:MForSWithThreeCUEs}), and is numerically verified in figure~\ref{fig:EIG} (B), with excellent concord in the bulk.


\subsubsection{Example 3: Arbitrary $L$ and two ``degenerate'' weights}
\label{sss:EIGExample3}

Let now $L$ be arbitrary, but the weights assume only two values --- let \smash{$L_{1}$} weights be equal to some \smash{$w_{1}$}, and the other \smash{$L_{2}$} weights to some \smash{$w_{2}$} (\smash{$L_{1} + L_{2} = L$}). The master equation can be transformed into a third--order polynomial equation,
\begin{equation}
\begin{split}\label{eq:MForSWithL1L2CUEs}
&M^{3} \left( R - L_{1} | w_{1} | - L_{2} | w_{2} | \right) \left( R + L_{1} | w_{1} | - L_{2} | w_{2} | \right) \cdot\\
&\cdot \left( R - L_{1} | w_{1} | + L_{2} | w_{2} | \right) \left( R + L_{1} | w_{1} | + L_{2} | w_{2} | \right) +\\
&+ 2 M^{2} \Big( L R^{4} - ( L + 1 ) \left( L_{1}^{2} | w_{1} |^{2} + L_{2}^{2} | w_{2} |^{2} \right) R^{2} +\\
&+ \left( L_{1}^{2} | w_{1} |^{2} - L_{2}^{2} | w_{2} |^{2} \right)^{2} \Big) +\\
&+ M \Big( \left( L^{2} + L_{1} L_{2} \right) R^{4} -\\
&- L \left( \left( 2 + L_{2} \right) L_{1}^{2} | w_{1} |^{2} + \left( 2 + L_{1} \right) L_{2}^{2} | w_{2} |^{2} \right) R^{2} +\\
&+ \left( L_{1}^{2} | w_{1} |^{2} - L_{2}^{2} | w_{2} |^{2} \right)^{2} \Big) +\\
&+ L L_{1} L_{2} \left( R^{2} - L_{1} | w_{1} |^{2} - L_{2} | w_{2} |^{2} \right) R^{2} = 0 .
\end{split}
\end{equation}
Remark that (\ref{eq:MForSWithL1L2CUEs}) reduces to a quadratic equation when one of the lengths \smash{$L_{1 , 2}$} equals $1$, and further to a linear equation when both lengths are $1$, which case has been discussed in paragraph~\ref{sss:EIGExample1}.

The values $M = 0$ or $M = - 1$ substituted into (\ref{eq:MForSWithL1L2CUEs}) lead to the conclusion that if \smash{$L_{1 , 2} \geq 2$}, then $\mathcal{D}$ is a disk of radius
\begin{equation}\label{eq:RExtForSWithL1L2CUEs}
R_{\textrm{ext}} = \sqrt{L_{1} | w_{1} |^{2} + L_{2} | w_{2} |^{2}} .
\end{equation}
However, when one of the lengths equals $1$, say \smash{$L_{1}$}, then $\mathcal{D}$ may happen to be an annulus of internal radius
\begin{equation}\label{eq:RIntForSWithL1L2CUEs}
R_{\textrm{int}} = \sqrt{\max \left( | w_{1} |^{2} - L_{2} | w_{2} |^{2} , 0 \right)} .
\end{equation}
These expressions agree with the general hypothesis stated at the end of paragraph~\ref{sss:EIGExample2}.

The radial mean spectral density stemming from (\ref{eq:RadialMeanSpectralDensityFromRotationallySymmetricNonHolomorphicMTransform}) and equation (\ref{eq:MForSWithL1L2CUEs}) is numerically tested in figure~\ref{fig:EIG} (C), showing perfect coincidence in the bulk.


\subsubsection{Example 4: Arbitrary $L$ and equal weights. A central limit theorem}
\label{sss:EIGExample4}

Let again $L$ be arbitrary, and this time all the weights be equal, \smash{$w_{l} \equiv w$}. The master equation turns into a linear one, which yields the non--holomorphic solution,
\begin{equation}\label{eq:MForSWithLCUEs}
M = - \frac{R_{\textrm{ext}}^{2} - R^{2}}{R_{\textrm{ext}}^{2} - \frac{R^{2}}{L}} ,
\end{equation}
where \smash{$R_{\textrm{ext}}$} is given by (\ref{eq:RExtForSWithLCUEs}). Let me also print the value of
\begin{equation}\label{eq:CForSWithLCUEs}
C = \frac{\left( 1 - \frac{1}{L} \right) \left( R_{\textrm{ext}}^{2} - R^{2} \right)}{\left( R_{\textrm{ext}}^{2} - \frac{R^{2}}{L} \right)^{2}} .
\end{equation}

The mean spectral domain $\mathcal{D}$ is always a disk with radius
\begin{equation}\label{eq:RExtForSWithLCUEs}
R_{\textrm{ext}} = \sqrt{L} | w | .
\end{equation}

The radial mean spectral density (\ref{eq:RadialMeanSpectralDensityFromRotationallySymmetricNonHolomorphicMTransform}), (\ref{eq:MForSWithLCUEs}),
\begin{equation}\label{eq:RhoForSWithLCUEs}
\rho^{\textrm{rad}}_{\mathbf{S}} ( R ) = \frac{2 R R_{\textrm{ext}}^{2} \left( 1 - \frac{1}{L} \right)}{\left( R_{\textrm{ext}}^{2} - \frac{R^{2}}{L} \right)^{2}} ,
\end{equation}
for \smash{$R \leq R_{\textrm{ext}}$}, and zero otherwise. It is checked numerically in figure~\ref{fig:EIG} (D), with perfect agreement in the bulk. This formula has been first derived in~\cite{HaagerupLarsen2000}, and then independently re--derived in~\cite{GorlichJarosz2004}.

It is interesting to consider the limit $L \to \infty$. It is clear from the above results that in order to arrive at a finite distribution, the weight $w$ must scale as \smash{$w \sim 1 / \sqrt{L}$}. Taking $L$ large, one finds that the mean spectral density (\ref{eq:RhoForSWithLCUEs}) becomes constant, \smash{$\rho_{\mathbf{S}} ( z , z^{*} ) = 1 / ( \pi R_{\textrm{ext}}^{2} )$}, inside a disk of radius \smash{$R_{\textrm{ext}}$} --- which is the ``Ginibre unitary ensemble'' (GinUE)~\cite{Ginibre1965,Girko19841985,TaoVu2007}. It is visible in figure~\ref{fig:EIG} (D). This is a first instance of the central limit theorem for unitary random matrices, here proven for free CUE's with equal weights; see paragraph~\ref{sss:EIGExample5} and subsection~\ref{ss:CentralLimitTheorem} for generalizations.


\subsubsection{Example 5: $L \to \infty$ and small weights. A central limit theorem}
\label{sss:EIGExample5}

A simple generalization of the central limit theorem discussed in the previous paragraph would be to take $L \to \infty$ in the sum of free CUE's with arbitrary weights, only assuming \smash{$| w_{l} | \ll 1$}. This may include situations such as \smash{$w_{l} = w ( i / L ) / \sqrt{L}$}, where $w ( x )$ is a function on $[ 0 , 1 ]$.

Here it is convenient to use the master equation in the form (\ref{eq:MainEquationForMForSWithCUEsVersion2Equation1})--(\ref{eq:MainEquationForMForSWithCUEsVersion2Equation3}). Since \smash{$| w_{l} | \ll 1$}, (\ref{eq:MainEquationForMForSWithCUEsVersion2Equation3}) implies that also \smash{$M_{l} \ll 1$}, if only we require a finite end result. Thus (\ref{eq:MainEquationForMForSWithCUEsVersion2Equation3}) becomes \smash{$M_{l} \approx - C | w_{l} |^{2}$}. Denoting
\begin{equation}\label{eq:CLTSigma}
\sigma^{2} \equiv \sum_{l = 1}^{L} | w_{l} |^{2} ,
\end{equation}
and supposing it to be finite and non--zero, the last equation substituted into (\ref{eq:MainEquationForMForSWithCUEsVersion2Equation1}) gives \smash{$- C \approx M / \sigma^{2}$}, upon which (\ref{eq:MainEquationForMForSWithCUEsVersion2Equation2}) finally yields the GinUE Green function \smash{$G \approx z^{*} / \sigma^{2}$}, \ie the constant mean spectral density \smash{$1 / ( \pi \sigma^{2} )$} inside a disk of radius \smash{$R_{\textrm{ext}} = \sigma$}.


\subsection{Central limit theorem}
\label{ss:CentralLimitTheorem}

Let me resume the discussion from paragraph~\ref{sss:EIGExample4}, and again investigate the sum $\mathbf{S}$ (\ref{eq:SDefinition}) with equal weights \smash{$w_{l} \equiv 1 / \sqrt{L}$}, in the limit $L \to \infty$, however now without any restriction on the probability distribution of the summands \smash{$\mathbf{U}_{l}$}, except that it is identical for all the matrices and that its first moment (drift) vanishes, \smash{$m_{1} = 0$}. I have already shown that if the \smash{$\mathbf{U}_{l}$}'s belong to the CUE class, the resulting mean spectral density is constant inside $C ( 0 , 1 )$ (GinUE with variance $1$). I will now prove that for an arbitrary ensemble of the \smash{$\mathbf{U}_{l}$}'s, the mean spectral density of $\mathbf{S}$ tends with increasing $L$ to a universal distribution (independent of the details of the unitary ensemble, but dependent only on its second moment \smash{$m_{2}$}), namely the constant density inside a certain ellipse (\ref{eq:CLTEllipse}).

To do that, I substitute into the system of equations (\ref{eq:QuaternionGreenFunctionaForUnitary})--(\ref{eq:QuaternionGreenFunctionbForUnitary}) and (\ref{eq:QuaternionAdditionLawcForSWithCUEs})--(\ref{eq:QuaternionAdditionLawdForSWithCUEs}), along with (\ref{eq:HolomorphicMTransformForUnitary}), the large--$L$ expansions
\begin{subequations}
\begin{align}
a &= a_{0} + \frac{1}{\sqrt{L}} a_{1} + \frac{1}{L} a_{2} + \ldots ,\label{eq:aLargeLExpansion}\\
b &= b_{0} + \frac{1}{\sqrt{L}} b_{1} + \frac{1}{L} b_{2} + \ldots ,\label{eq:bLargeLExpansion}
\end{align}
\end{subequations}
and solve them order by order. Recall that \smash{$a = G_{\mathbf{S}} ( z , z^{*} )$} and \smash{$| b |^{2} = C_{\mathbf{S}} ( z , z^{*} )$}.

At the leading order, one discovers
\begin{subequations}
\begin{align}
a_{0} &= \frac{z^{*} - m_{2}^{*} z}{1 - \left| m_{2} \right|^{2}} ,\label{eq:CLTa0}\\
\left| b_{0} \right|^{2} &= 1 + \frac{- | z |^{2} \left( 1 + \left| m_{2} \right|^{2} \right) + m_{2} ( z^{*} )^{2} + m_{2}^{*} z^{2}}{\left( 1 - \left| m_{2} \right|^{2} \right)^{2}} .\label{eq:CLTb0}
\end{align}
\end{subequations}
Hence, the borderline of the mean spectral domain $\mathcal{D}$ (given by \smash{$| b_{0} |^{2} = 0$}) is the ellipse
\begin{equation}
\begin{split}\label{eq:CLTEllipse}
&\left( \left( 1 - \re m_{2} \right)^{2} + ( \im m_{2} )^{2} \right) x^{2} - 4 ( \im m_{2} ) x y +\\
&+ \left( \left( 1 + \re m_{2} \right)^{2} + ( \im m_{2} )^{2} \right) y^{2} = \left( 1 - \left| m_{2} \right|^{2} \right)^{2} ,
\end{split}
\end{equation}
which has the semi--axes \smash{$( 1 \pm | m_{2} | )$}, the angle \smash{$\varphi \in [ - \frac{\pi}{2} , \frac{\pi}{2} )$} from the $x$--axis to the major axis of the ellipse is \smash{$\varphi = \frac{1}{2} \textrm{Arg} ( m_{2} )$} (with the convention that the principal argument lies in $[ - \pi , \pi )$), while its area reads \smash{$\pi ( 1 - | m_{2} |^{2} )$}. Furthermore, the mean spectral density (\ref{eq:CLTa0}), (\ref{eq:MeanSpectralDensityFromNonHolomorphicGreenFunction}) is constant, \smash{$\rho_{\mathbf{S}} ( z , z^{*} ) = 1 / ( \pi ( 1 - | m_{2} |^{2} ) )$}, within this ellipse. This is the central limit theorem for free identically--distributed zero--drift unitary random matrices.

I have also derived the next--to--leading--order expressions,
\begin{subequations}
\begin{align}
a_{1} &= \frac{m_{3} m_{2}^{*} \left( z^{*} - m_{2}^{*} z \right)^{2} - m_{3}^{*} \left( z - m_{2} z^{*} \right)^{2}}{\left( 1 - \left| m_{2} \right|^{2} \right)^{3}} ,\label{eq:CLTa1}\\
b_{1} b_{0}^{*} &= \re \left( \frac{m_{3} \left( z^{*} - m_{2}^{*} z \right)^{2} \left( \left( 1 + \left| m_{2} \right|^{2} \right) z^{*} - 2 m_{2}^{*} z \right)}{\left( 1 - \left| m_{2} \right|^{2} \right)^{4}} \right) .\label{eq:CLTb1}
\end{align}
\end{subequations}
(Note, \smash{$| b |^{2} = | b_{0} |^{2} + \frac{1}{\sqrt{L}} 2 b_{1} b_{0}^{*} + \ldots$}.) They are proportional to the third moment \smash{$m_{3}$}.


\subsection{Finite--size effects}
\label{ss:FiniteSizeEffects}


\subsubsection{The $\erfc$ form--factor}
\label{sss:TheErfcFormFactor}

\begin{figure*}[t]
\includegraphics[width=\columnwidth]{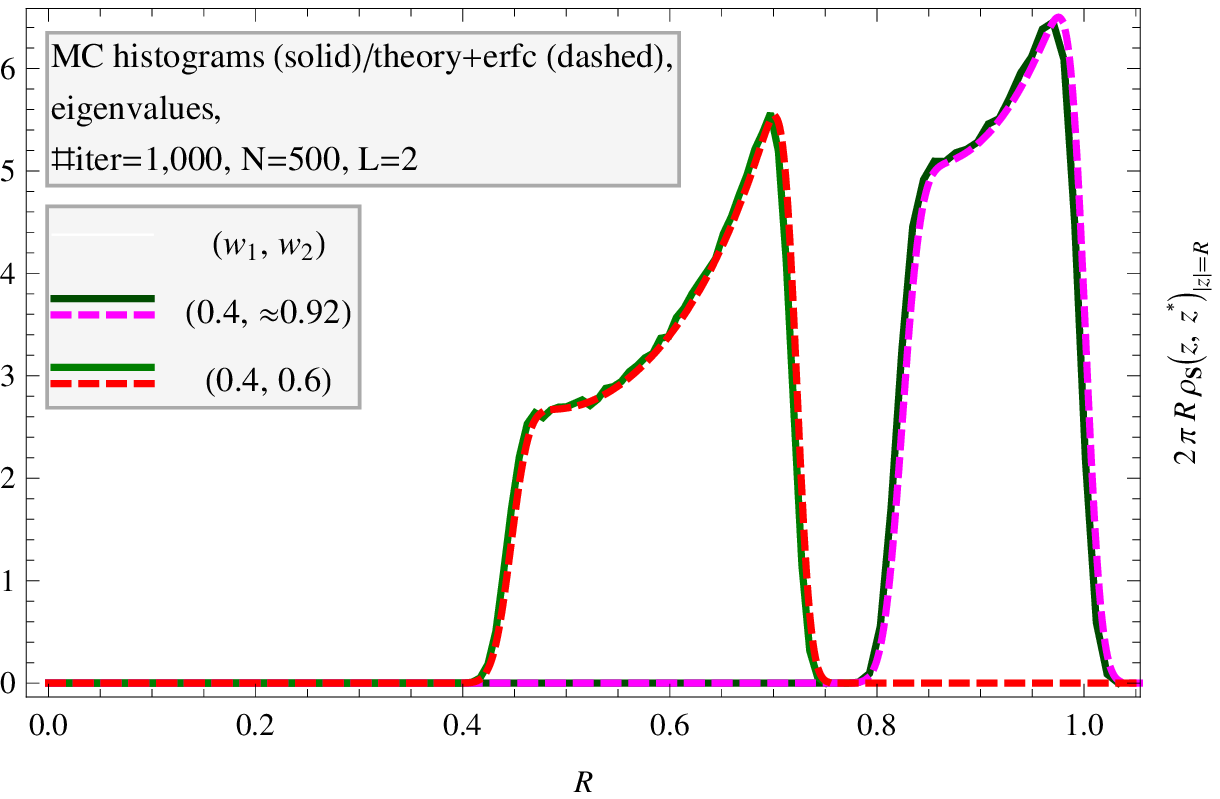}
\includegraphics[width=\columnwidth]{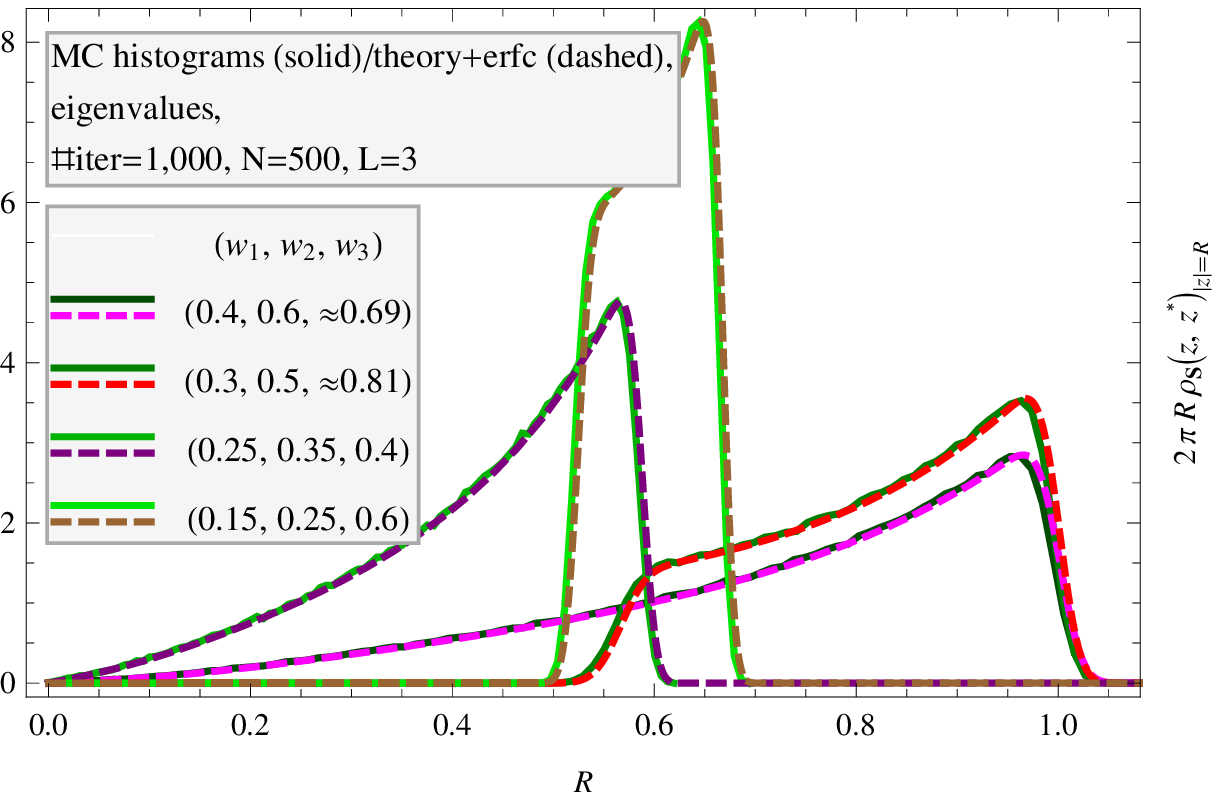}
\includegraphics[width=\columnwidth]{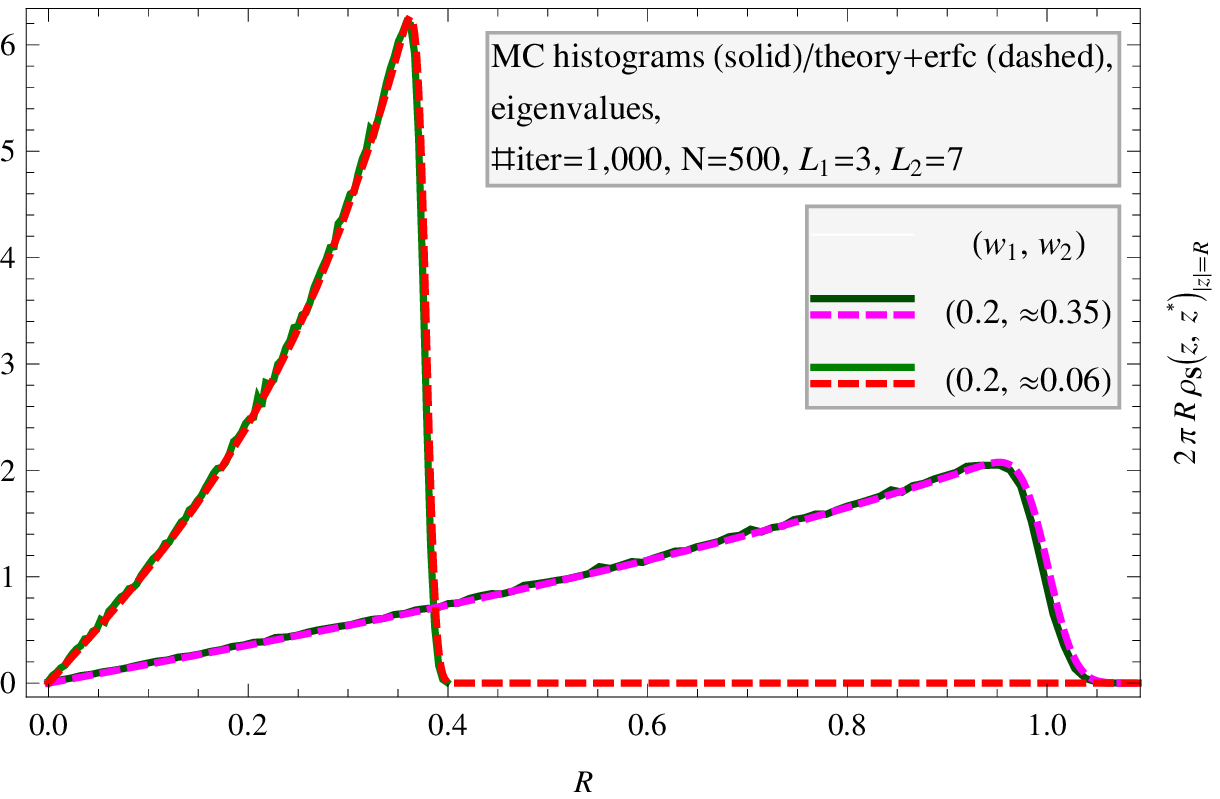}
\includegraphics[width=\columnwidth]{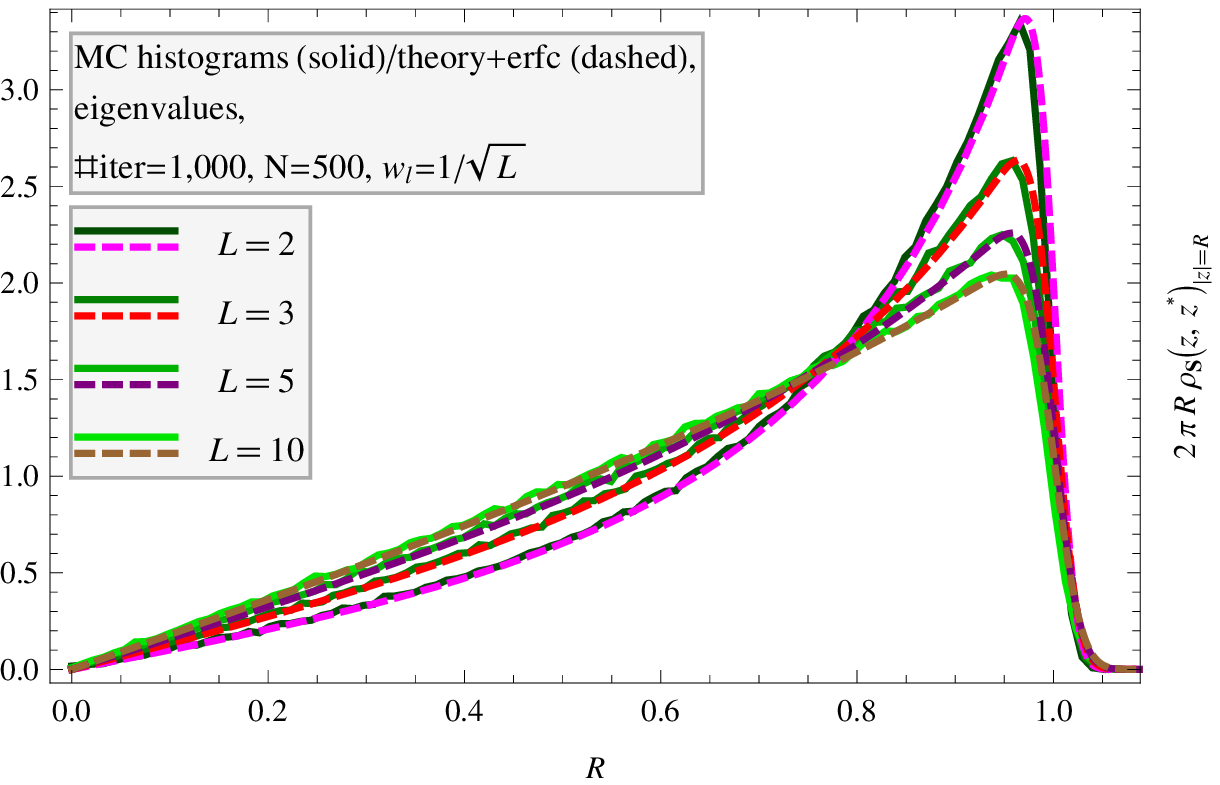}
\caption{The same numerical histograms as in figure~\ref{fig:EIG}, checked against the finite--$N$ conjecture \smash{$\rho^{\textrm{rad}}_{\mathbf{S}} ( R ) f_{N , q , R_{\textrm{ext}} , + 1} ( R )$} (when $\mathcal{D}$ is a disk) or \smash{$\rho^{\textrm{rad}}_{\mathbf{S}} ( R ) f_{N , q_{1} , R_{\textrm{ext}} , + 1} ( R ) f_{N , q_{2} , R_{\textrm{int}} , - 1} ( R )$} (when $\mathcal{D}$ is an annulus) (\ref{eq:FiniteSizeBorderlineFactor}), where the least--squares values of the parameter(s) are:\\
A: \smash{$( q_{1} , q_{2} ) \approx$} (1) $( 2.61 , 2.22 )$, (2) $( 2.98 , 2.33 )$.\\
B: (1) $q \approx 1.82$, (2) \smash{$( q_{1} , q_{2} ) \approx ( 2.01 , 1.56 )$}, (3) $q \approx 3.03$, (4) \smash{$( q_{1} , q_{2} ) \approx ( 3.55 , 3.07 )$}.\\
C: $q \approx$ (1) $1.49$, (2) $4.27$.\\
D: $q \approx$ (1) $2.04$, (2) $1.73$, (3) $1.57$, (4) $1.47$.}
\label{fig:EIGErfc}
\end{figure*}

Random matrices exhibit freeness only in the limit of large matrix dimensions~\cite{VoiculescuDykemaNica1992}, and consequently, all the above results hold only for $N \to \infty$, so in particular are incapable of capturing the finite--$N$ behavior of the mean spectral density close to the borderline $\partial \mathcal{D}$ (\emph{cf.} the strong dumping visible in figure~\ref{fig:EIG}).

However, this behavior has been described for a number of non--Hermitian random matrix models ($N \times N$) which display rotationally--symmetric mean spectrum and for which $\mathcal{D}$ is a disk \smash{$R = R_{\textrm{ext}}$} --- by a simple form--factor, \smash{$\frac{1}{2} \erfc ( q ( R - R_{\textrm{ext}} ) \sqrt{N} )$}, where \smash{$\erfc ( x ) \equiv \frac{2}{\sqrt{\pi}} \int_{x}^{\infty} \dd t \exp ( - t^{2} )$} is the complementary error function, while $q$ depends on the particular model. These models are:
\begin{itemize}
\item GinUE~\cite{ForresterHonner1999,Kanzieper2005} (see also~\cite{KhoruzhenkoSommers2009}); $q$ has been derived explicitly.
\item $\mathbf{P}$ (\ref{eq:PDefinition}) with $K = 2$ and arbitrary \smash{$R_{2}$}~\cite{KanzieperSingh2010}; $q$ has also been computed.
\item $\mathbf{P}$ with arbitrary $K$ and arbitrary \smash{$R_{k}$}'s~\cite{BurdaJaroszLivanNowakSwiech20102011}; the above form--factor has been conjectured and verified numerically, with $q$ treated as a parameter to be fitted by comparison with experimental data, as its exact form is yet unknown.
\item The ``time--lagged covariance estimator'' \smash{$\mathbf{c} \propto \mathbf{A} \mathbf{D} \mathbf{A}^{\dagger}$}, where $\mathbf{A}$ is a rectangular Gaussian random matrix (\ref{eq:RectangularGGMeasure}), and \smash{$D_{a b} \equiv \delta_{a + 1 , b}$}~\cite{Jarosz2010-01}; $q$ has been again treated as an adjustable parameter and remains to be derived.
\end{itemize}


\subsubsection{The $\erfc$ form--factor for weighted sums of free CUE random matrices}
\label{sss:TheErfcFormFactorForWeightedSumsOfFreeUnitaryRandomMatrices}

Prompted by this performance of the $\erfc$ form--factor, I propose the following conjecture for the finite--$N$ form of the mean spectral density of any weighted sum $\mathbf{S}$ (\ref{eq:SDefinition}) of CUE random matrices: The radial part of the $N \to \infty$ formula for the density, \smash{$\rho^{\textrm{rad}}_{\mathbf{S}} ( R )$} (\ref{eq:RadialMeanSpectralDensityFromRotationallySymmetricNonHolomorphicMTransform}), should be multiplied by the factor of
\begin{equation}\label{eq:FiniteSizeBorderlineFactor}
f_{N , q_{\textrm{b}} , R_{\textrm{b}} , s_{\textrm{b}}} ( R ) \equiv \frac{1}{2} \erfc \left( q_{\textrm{b}} s_{\textrm{b}} \left( R - R_{\textrm{b}} \right) \sqrt{N} \right) ,
\end{equation}
for each centered circle \smash{$R = R_{\textrm{b}}$} constituting a connected part of the borderline $\partial \mathcal{D}$ (there can be only either one or two such circles, \ie $\mathcal{D}$ is either a disk or an annulus, as we know from the single ring theorem; recall Remark 5 in paragraph~\ref{sss:EIGTheMasterEquation}), where the sign \smash{$s_{\textrm{b}}$} is $+ 1$ for the external borderline and $- 1$ for the internal borderline, while \smash{$q_{\textrm{b}}$} is at the moment an adjustable parameter (one for each circle) to be found by fitting to experimental data, and eventually to be calculated.

Figure~\ref{fig:EIGErfc} shows numerical tests of this conjecture for the same cases as in figure~\ref{fig:EIG}; the proposal performs very well on all the examples.


\subsubsection{The $\erfc$ conjecture}
\label{sss:TheErfcConjecture}

Since the $\erfc$ form--factor proves to perfectly reproduce the Monte--Carlo data for the four very different matrix models described in paragraphs~\ref{sss:TheErfcFormFactor} and~\ref{sss:TheErfcFormFactorForWeightedSumsOfFreeUnitaryRandomMatrices}, I put forth a conjecture that (\ref{eq:FiniteSizeBorderlineFactor}) is valid for any non--Hermitian random matrix model whose mean spectrum possesses rotational symmetry around zero.

Besides proving this hypothesis, another challenge is to express in each case the parameter(s) \smash{$q_{\textrm{b}}$} through the parameters of the given model.

In particular, it should work for $\mathbf{W}$, provided that the \smash{$\mathbf{U}_{l}$}'s belong to the CUE.


\section{Summing free unitary random matrices --- the singular values}
\label{s:SVSummingFreeUnitaryRandomMatrices}

In this section, I will continue an analysis of a weighted sum of free unitary random matrices $\mathbf{S}$ (\ref{eq:SDefinition}) in the large--$N$ limit --- focusing on the singular values.


\subsection{Conjecture about rotationally--symmetric spectra}
\label{ss:ConjectureAboutRotationallySymmetricSpectra}

As noted in paragraph~\ref{sss:EIGTheMasterEquation}, Remark 3, the mean spectral density of a weighted sum of free CUE's (but not necessarily more general unitary ensembles!) is rotationally--symmetric around zero. In recent papers~\cite{BurdaJaroszLivanNowakSwiech20102011,Jarosz2010-01}, the following conjecture has been claimed, which relates the mean eigenvalues and singular values of any non--Hermitian random matrix $\mathbf{X}$ with the mentioned symmetry property:
\begin{description}
\item[Step 1:] The assumed symmetry of $\mathbf{X}$ can be restated as the rotational symmetry around zero of the non--holomorphic $M$--transform (\ref{eq:NonHolomorphicMTransformDefinition}), \smash{$M_{\mathbf{X}} ( z , \overline{z} ) = \mathfrak{M}_{\mathbf{X}} ( R^{2} )$}. This allows to define its functional inverse,
    \begin{equation}\label{eq:RotationallySymmetricNonHolomorphicNTransformDefinition}
    \mathfrak{M}_{\mathbf{X}} ( \mathfrak{N}_{\mathbf{X}} ( z ) ) = z ,
    \end{equation}
    called the ``rotationally--symmetric non--holomorphic $N$--transform.''
\item[Step 2:] The random matrix \smash{$\mathbf{X}^{\dagger} \mathbf{X}$} is Hermitian, thus one can always compute its holomorphic $M$--transform (\ref{eq:HolomorphicMTransformDefinition}), \smash{$M_{\mathbf{X}^{\dagger} \mathbf{X}} ( z )$}, and the ``holomorphic $N$--transform'' being its functional inverse,
    \begin{equation}\label{eq:HolomorphicNTransformDefinition}
    M_{\mathbf{X}^{\dagger} \mathbf{X}} ( N_{\mathbf{X}^{\dagger} \mathbf{X}} ( z ) ) = z .
    \end{equation}
\item[Step 3:] The conjecture states that the two above $N$--transforms remain in the following relationship,
    \begin{equation}\label{eq:ConjectureAboutRotationallySymmetricSpectra}
    N_{\mathbf{X}^{\dagger} \mathbf{X}} ( z ) = \frac{z + 1}{z} \mathfrak{N}_{\mathbf{X}} ( z ) .
    \end{equation}
\end{description}


\subsection{Summing free CUE random matrices}
\label{ss:SVSummingFreeCUERandomMatrices}


\subsubsection{The master equation}
\label{sss:SVTheMasterEquation}

It is straightforward to apply the hypothesis (\ref{eq:ConjectureAboutRotationallySymmetricSpectra}) to the master equation (\ref{eq:MainEquationForMForSWithCUEs}); thus the master equation for the holomorphic $M$--transform, or better, the Green function \smash{$G \equiv G_{\mathbf{S}^{\dagger} \mathbf{S}} ( z )$}, for free CUE's, reads
\begin{equation}\label{eq:SVMainEquationForGForSWithCUEs}
L - 2 + 2 z G = \sum_{l = 1}^{L} s_{l} \sqrt{1 + 4 \left| w_{l} \right|^{2} z G^{2}} .
\end{equation}

\begin{description}
\item[Remark 1:] (\ref{eq:SVMainEquationForGForSWithCUEs}) may be recast as a set of equations analogous to (\ref{eq:MainEquationForMForSWithCUEsVersion2Equation1})--(\ref{eq:MainEquationForMForSWithCUEsVersion2Equation3}), with just one equation different,
    \begin{equation}\label{eq:SVMainEquationForGForSWithCUEsVersion2Equation1}
    z G - 1 = \sum_{l = 1}^{L} M_{l} ,
    \end{equation}
    where
    \begin{subequations}
    \begin{align}
    - z G^{2} &= C ,\label{eq:SVMainEquationForGForSWithCUEsVersion2Equation2}\\
    - \frac{M_{l} ( M_{l} + 1 )}{\left| w_{l} \right|^{2}} &= C , \quad l = 1 , 2 , \ldots , L .\label{eq:SVMainEquationForGForSWithCUEsVersion2Equation3}
    \end{align}
    \end{subequations}
\item[Remark 2:] Recall that, as for any Hermitian random matrix, the Green function must behave at complex infinity $z \to \infty$ as $G \sim 1 / z$. This will help us choose the proper solution of the master equation.
\item[Remark 3:] It is a known fact~\cite{Zee1996,JanikNowakPappZahed1997-02} that for any Hermitian random matrix $\mathbf{H}$, the end--points \smash{$x_{\star}$} of the support of its mean spectral density, \smash{$\rho_{\mathbf{H}} ( x )$}, are the branch points of the Green function, \ie \smash{$\dd G_{\mathbf{H}} ( z ) / \dd z |_{z = x_{\star}} = \infty$}. Applied to the master equation (\ref{eq:SVMainEquationForGForSWithCUEs}), this condition reads
    \begin{equation}\label{eq:EndpointsOfTheSupportOfTheMeanSpectralDenistyOfSDaggerS}
    \frac{1}{2} = G_{\star} \sum_{l = 1}^{L} \frac{s_{l} \left| w_{l} \right|^{2}}{\sqrt{1 + 4 \left| w_{l} \right|^{2} x_{\star} G_{\star}^{2}}} .
    \end{equation}
    Thus the end--points are found by solving the set of equations: (\ref{eq:SVMainEquationForGForSWithCUEs}) (with \smash{$( z , G ) = ( x_{\star} , G_{\star} )$}) and (\ref{eq:EndpointsOfTheSupportOfTheMeanSpectralDenistyOfSDaggerS}), for \smash{$x_{\star} \geq 0$}.
\end{description}

I will now revisit the five examples from subsection~\ref{ss:EIGSummingFreeCUERandomMatrices}, calculating the mean singular values. One will observe that in each case, the polynomial equation for $G$ will have an order greater by $1$ from the corresponding equation for the mean eigenvalues.


\subsubsection{Example 1: $L = 2$}
\label{sss:SVExample1}

\begin{figure*}[t]
\includegraphics[width=\columnwidth]{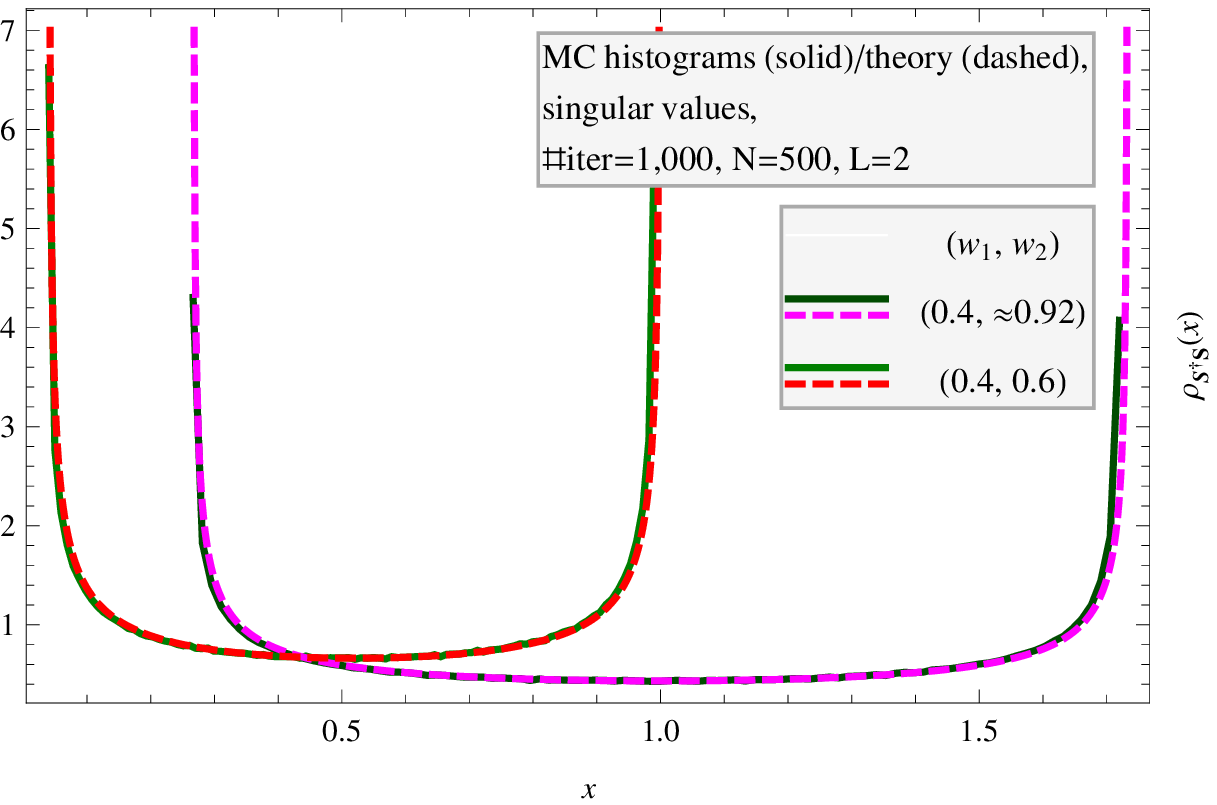}
\includegraphics[width=\columnwidth]{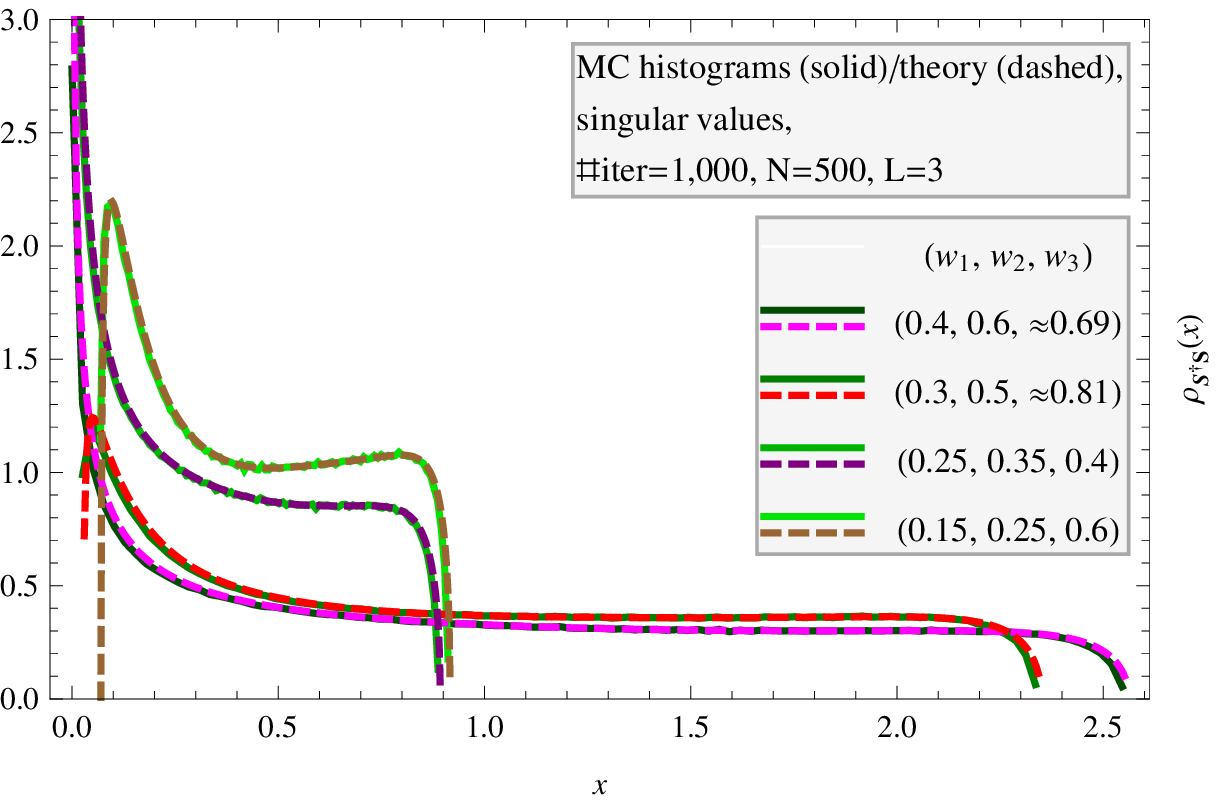}
\includegraphics[width=\columnwidth]{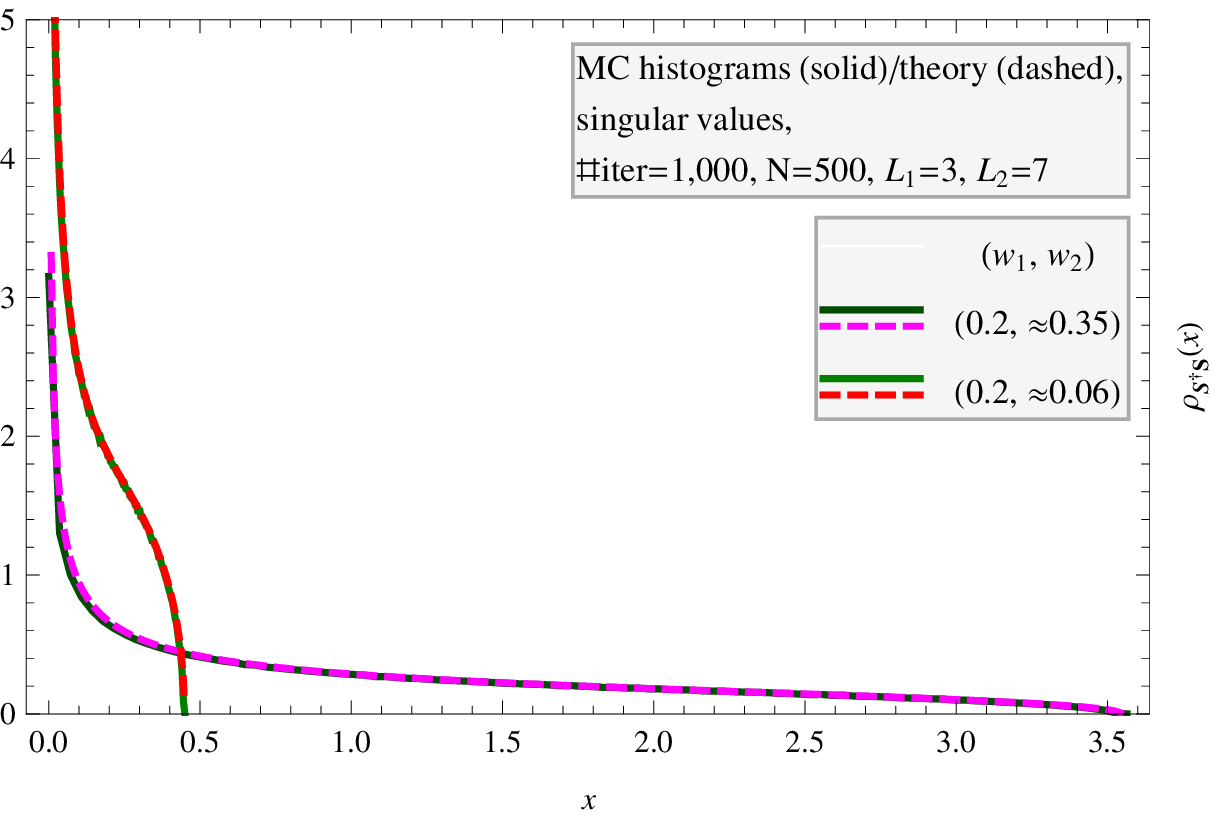}
\includegraphics[width=\columnwidth]{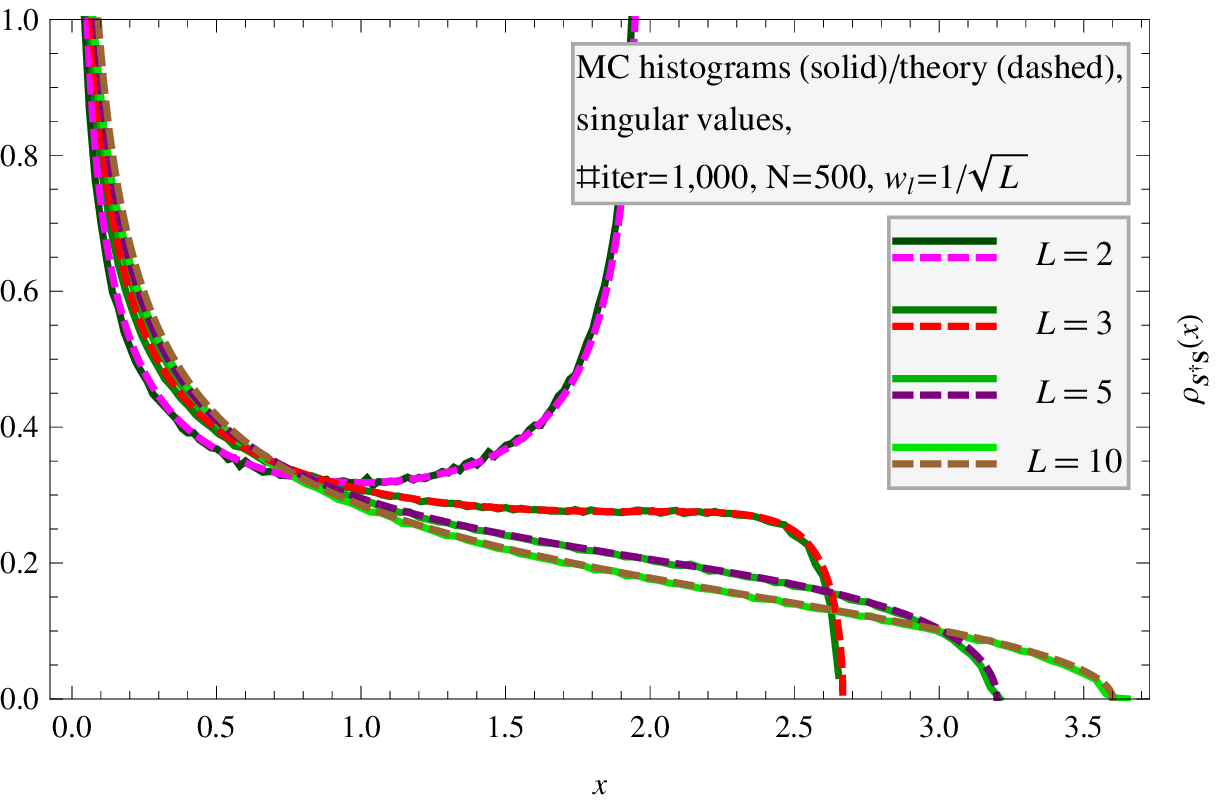}
\caption{The mean density of the singular values, \smash{$\rho_{\mathbf{S}^{\dagger} \mathbf{S}} ( x )$}, for the same values of $L$ and the weights as in figure~\ref{fig:EIG}.}
\label{fig:SV}
\end{figure*}

Consider a sum of $L = 2$ free CUE matrices, with arbitrary weights \smash{$w_{1 , 2}$}. The master equation becomes quadratic, and its solution with the proper large--$z$ asymptotics reads
\begin{equation}\label{eq:SVGForSWithTwoCUEs}
G_{\mathbf{S}^{\dagger} \mathbf{S}} ( z ) = \frac{1}{\sqrt{z - \left( \left| w_{1} \right| - \left| w_{2} \right| \right)^{2}} \sqrt{z - \left( \left| w_{1} \right| + \left| w_{2} \right| \right)^{2}}} .
\end{equation}

The mean spectral density of \smash{$\mathbf{S}^{\dagger} \mathbf{S}$} follows from (\ref{eq:MeanSpectralDensityFromHolomorphicGreenFunction}) applied to (\ref{eq:SVGForSWithTwoCUEs}),
\begin{equation}\label{eq:SVRhoForSWithTwoCUEs}
\rho_{\mathbf{S}^{\dagger} \mathbf{S}} ( x ) = \frac{1}{\pi \sqrt{\left( \left( \left| w_{1} \right| - \left| w_{2} \right| \right)^{2} - x \right) \left( x - \left( \left| w_{1} \right| + \left| w_{2} \right| \right)^{2} \right)}} ,
\end{equation}
for \smash{$( | w_{1} | - | w_{2} | )^{2} < x < ( | w_{1} | + | w_{2} | )^{2}$}, and zero otherwise. This is numerically checked in figure~\ref{fig:SV} (A), with perfect concord. Note that changing the argument \smash{$x = ( | w_{1} | - | w_{2} | )^{2} + 4 | w_{1} | | w_{2} | x^{\prime}$}, where \smash{$0 < x^{\prime} < 1$}, the density (\ref{eq:SVRhoForSWithTwoCUEs}) becomes \smash{$\rho_{\mathbf{S}^{\dagger} \mathbf{S}}^{\prime} ( x^{\prime} ) = 1 / ( \pi \sqrt{x^{\prime} ( 1 - x^{\prime} )} )$}, which is the ``arcsine distribution.''


\subsubsection{Example 2: $L = 3$}
\label{sss:SVExample2}

For a sum of $L = 3$ free CUE matrices, with arbitrary weights \smash{$w_{1 , 2 , 3}$}, the master equation turns into a fifth--order polynomial equation,
\begin{equation}
\begin{split}\label{eq:SVGForSWithThreeCUEs}
&G^{5} z \left( z - v_{0}^{2} \right) \left( z - v_{1}^{2} \right) \left( z - v_{2}^{2} \right) \left( z - v_{3}^{2} \right) +\\
&+ 4 G^{4} z \Big( z^{3} - 3 \mu_{(  1  )} z^{2} + \left( 3 \mu_{(  2  )} + 2 \mu_{(  1 , 1  )} \right) z -\\
&- \mu_{(  3  )} + \mu_{(  2 , 1  )} - 10 \mu_{(  1 , 1 , 1  )} \Big) +\\
&+ 2 G^{3} \left( 2 z^{3} - 5 \mu_{(  1  )} z^{2} + 4 \mu_{(  2  )} z + V_{1} V_{2} V_{3} \right) -\\
&- 2 G^{2} \left( z^{2} + v_{0} v_{1} v_{2} v_{3} \right) +\\
&+ G \left( -5 z + 2 \mu_{(  1  )} \right) - 2 = 0 ,
\end{split}
\end{equation}
where recall paragraph~\ref{sss:EIGExample2} for notation.

The mean density of the singular values is obtained by solving numerically (\ref{eq:SVGForSWithThreeCUEs}) at $z = \lambda \pm \ii \epsilon$, for small $\epsilon$ (\ref{eq:MeanSpectralDensityFromHolomorphicGreenFunction}), which is compared with Monte--Carlo simulations in figure~\ref{fig:SV} (B), discovering perfect agreement.


\subsubsection{Example 3: Arbitrary $L$ and two ``degenerate'' weights}
\label{sss:SVExample3}

For arbitrary \smash{$L = L_{1} + L_{2}$}, and \smash{$L_{1}$} weights \smash{$w_{1}$} and \smash{$L_{2}$} weights \smash{$w_{2}$}, the master equation can be transformed into a fourth--order polynomial equation,
\begin{equation}
\begin{split}\label{eq:SVGForSWithL1L2CUEs}
&G^{4} z^{2} \left( z - \left( L_{1} | w_{1} | + L_{2} | w_{2} | \right)^{2} \right) \left( z - \left( L_{1} | w_{1} | - L_{2} | w_{2} | \right)^{2} \right) +\\
&+ 2 G^{3} z^{2} ( L - 2 ) \left( z - L_{1}^{2} | w_{1} |^{2} - L_{2}^{2} | w_{2} |^{2} \right) +\\
&+ G^{2} z \Big( \left( L^{2} - 6 L + 6 + L_{1} L_{2} \right) z +\\
&+ L_{1}^{2} \left( - L L_{2} + 2 L - 2 \right) | w_{1} |^{2} +\\
&+ L_{2}^{2} \left( - L L_{1} + 2 L - 2 \right) | w_{2} |^{2} \Big) +\\
&+ G z ( L - 2 ) \left( - 2 L + 2 + L_{1} L_{2} \right) -\\
&- ( L - 1 ) \left( L_{1} - 1 \right) \left( L_{2} - 1 \right) = 0 .
\end{split}
\end{equation}

The mean density of the singular values numerically computed from (\ref{eq:SVGForSWithL1L2CUEs}), (\ref{eq:MeanSpectralDensityFromHolomorphicGreenFunction}) and Monte--Carlo simulated is shown in figure~\ref{fig:SV} (C), finding excellent agreement between the two.


\subsubsection{Example 4: Arbitrary $L$ and equal weights. A central limit theorem}
\label{sss:SVExample4}

For arbitrary $L$ and all the weights equal to some $w$, the master equation becomes quadratic, yielding one solution with the proper asymptotics at infinity,
\begin{equation}\label{eq:SVGForSWithLCUEs}
G_{\mathbf{S}^{\dagger} \mathbf{S}} ( z ) = \frac{\frac{1}{2} - \frac{1}{L} - \frac{\sqrt{\frac{z}{4} - \left( 1 - \frac{1}{L} \right) R_{\textrm{ext}}^{2}}}{\sqrt{z}}}{R_{\textrm{ext}}^{2} - \frac{z}{L}} ,
\end{equation}
where \smash{$R_{\textrm{ext}}$} is given by (\ref{eq:RExtForSWithLCUEs})

Therefore, the mean density of the singular values (\ref{eq:MeanSpectralDensityFromHolomorphicGreenFunction}),
\begin{equation}\label{eq:SVRhoForSWithLCUEs}
\rho_{\mathbf{S}^{\dagger} \mathbf{S}} ( x ) = \frac{\sqrt{\left( 1 - \frac{1}{L} \right) R_{\textrm{ext}}^{2} \frac{1}{x} - \frac{1}{4} }}{R_{\textrm{ext}}^{2} - \frac{x}{L}} ,
\end{equation}
for \smash{$x \in [ 0 , 4 ( 1 - \frac{1}{L} ) R_{\textrm{ext}}^{2} ]$}, and zero otherwise. This is the known ``Kesten distribution''~\cite{Kesten1959}, and has been derived in~\cite{HaagerupLarsen2000}. It is successfully verified numerically in figure~\ref{fig:SV} (D).

Taking the limit $L \to \infty$ with \smash{$R_{\textrm{ext}}$} assumed finite (\ie \smash{$w \sim 1 / \sqrt{L}$}) leads of course to the Mar\v{c}enko--Pastur distribution~\cite{MarcenkoPastur1967}, \smash{$\rho_{\mathbf{S}^{\dagger} \mathbf{S}} ( x ) = \frac{1}{\pi R_{\textrm{ext}}} ( \frac{1}{x} - \frac{1}{4 R_{\textrm{ext}}^{2}} )^{1 / 2}$}, for \smash{$x \in \left[ 0 , 4 R_{\textrm{ext}}^{2} \right]$}, and zero otherwise.


\subsubsection{Example 5: $L \to \infty$ and small weights. A central limit theorem}
\label{sss:SVExample5}

For $L \to \infty$, in the presence of arbitrary weights such that \smash{$| w_{l} | \ll 1$}, an analogous procedure as in paragraph~\ref{sss:SVExample5} may be applied to the master equation in the form (\ref{eq:SVMainEquationForGForSWithCUEsVersion2Equation1})--(\ref{eq:SVMainEquationForGForSWithCUEsVersion2Equation3}); it gives a quadratic equation for the Green function, whose solution with the correct asymptotic behavior at infinite $z$ is
\begin{equation}\label{eq:SVGForCLT}
G_{\mathbf{S}^{\dagger} \mathbf{S}} ( z ) = \frac{1}{\sigma^{2}} \left( \frac{1}{2} - \frac{\sqrt{\frac{z}{4} - \sigma^{2}}}{\sqrt{z}} \right) ,
\end{equation}
where $\sigma$ is defined in (\ref{eq:CLTSigma}), and assumed finite and non--zero. This is again the Mar\v{c}enko--Pastur distribution, with variance (\ref{eq:CLTSigma}).


\section{Conclusions}
\label{s:Conclusions}


\subsection{Summary}
\label{ss:Summary}


\subsubsection{Main results}
\label{sss:MainResults}

In this article, I analyzed in the thermodynamic limit (\ref{eq:ThermodynamicLimit}) the weighted sum $\mathbf{S}$ (\ref{eq:SDefinition}) of independent unitary random matrices. The main results:
\begin{itemize}
\item The ``master equations'' (\ref{eq:MainEquationForMForSWithCUEs}) and (\ref{eq:SVMainEquationForGForSWithCUEs}), which yield the mean densities of the eigenvalues and singular values of $\mathbf{S}$, respectively, in case of all the \smash{$\mathbf{U}_{l}$}'s belonging to the CUE, but for arbitrary $L$ and weights.
\item These master equations are either solved or transformed into a polynomial form, and numerically verified, in four cases: (1) $L = 2$ and arbitrary weights, (2) $L = 3$ and arbitrary weights, (3) any $L$ and two ``degenerate'' weights, (4) any $L$ and equal weights. The resulting polynomial equations have orders: For the eigenvalues: (1) $1$, (2) $4$, (3) $3$, (4) $1$. For the singular values: (1) $2$, (2) $5$, (3) $4$, (4) $2$.
\item Two central limit theorems: (1) For the \smash{$\mathbf{U}_{l}$}'s being independent CUE matrices, and for arbitrary but small weights. The limiting distribution is the GinUE with variance (\ref{eq:CLTSigma}). (2) More generally, for the \smash{$\mathbf{U}_{l}$}'s being independent identically--distributed unitary random matrices of any probability distribution with first moment (drift) zero, and for equal weights, \smash{$w_{l} = 1 / \sqrt{L}$}. The limiting distribution is an elliptic modification of the GinUE (\ref{eq:CLTa0})--(\ref{eq:CLTb0}); I have also derived its first sub--leading correction (\ref{eq:CLTa1})--(\ref{eq:CLTb1}).
\item The conjecture about and numerical tests of the $\erfc$ form--factor (\ref{eq:FiniteSizeBorderlineFactor}).
\end{itemize}

These results, beyond being mathematically interesting, find applications \eg in quantum entanglement theory and theory of random walks on regular trees.


\subsubsection{Method}
\label{sss:Method}

Another goal of this article was to advertise quaternion free probability calculus (in particular, the quaternion addition law (\ref{eq:QuaternionAdditionLaw})), as a conceptually simple, purely algebraic and efficient tool to compute mean spectral densities of sums of free non--Hermitian random matrices.


\subsection{Open problems}
\label{ss:OpenProblems}


\subsubsection{Solve the model $\mathbf{W}$}
\label{sss:SolveTheModelW}

In a forthcoming publication, I plan to repeat the considerations of sections~\ref{s:EIGSummingFreeUnitaryRandomMatrices} (the mean spectral density) and~\ref{s:SVSummingFreeUnitaryRandomMatrices} (the mean singular values density) for the model $\mathbf{W}$ (\ref{eq:WDefinition}), with the supposition that the \smash{$\mathbf{U}_{l}$}'s belong to the CUE, and in the thermodynamic limit (\ref{eq:ThermodynamicLimit}). The outline of the procedure:
\begin{description}
\item[Step 1:] Consider first the singular values of $\mathbf{W}$, \ie the model \smash{$\mathbf{W}^{\dagger} \mathbf{W} = \mathbf{P}^{\dagger} \mathbf{S}^{\dagger} \mathbf{S} \mathbf{P}$}. Through cyclic shifts, it is easy to relate this matrix to the product of two Hermitian random matrices, \smash{$\mathbf{H}_{1} \equiv \mathbf{S}^{\dagger} \mathbf{S}$} and \smash{$\mathbf{H}_{2} \equiv \mathbf{P}^{\dagger} \mathbf{P}$}.
\item[Step 2:] The holomorphic $M$--transforms (\ref{eq:HolomorphicMTransformDefinition}) of both these matrices are known: \smash{$M_{\mathbf{H}_{1}} ( z )$} is given by equation (\ref{eq:SVMainEquationForGForSWithCUEs}), while \smash{$M_{\mathbf{H}_{2}} ( z )$} has been found in~\cite{BurdaJaroszLivanNowakSwiech20102011} to obey a polynomial equation of order $( K + 1 )$. Invert them functionally to obtain the respective holomorphic $N$--transforms (\ref{eq:HolomorphicNTransformDefinition}).
\item[Step 3:] Use the ``multiplication algorithm,'' well--known in free probability theory, valid for free \smash{$\mathbf{H}_{1 , 2}$}, \smash{$N_{\mathbf{H}_{1} \mathbf{H}_{2}} ( z ) = \frac{z}{1 + z} N_{\mathbf{H}_{1}} ( z ) N_{\mathbf{H}_{2}} ( z )$}.
\item[Step 4:] Invert functionally the result to get the holomorphic $M$--transform \smash{$M_{\mathbf{W}^{\dagger} \mathbf{W}} ( z )$}, which contains the information about the mean density of the singular values of $\mathbf{W}$.
\item[Step 5:] Since the mean spectral density of $\mathbf{W}$ must be rotationally--symmetric around zero (to be checked \emph{a posteriori}), the conjecture (\ref{eq:ConjectureAboutRotationallySymmetricSpectra}) finally yields the non--holomorphic $M$--transform \smash{$M_{\mathbf{W}} ( z , z^{*} )$} (\ref{eq:NonHolomorphicMTransformDefinition}), and consequently the mean spectral density of $\mathbf{W}$ (\ref{eq:RadialMeanSpectralDensityFromRotationallySymmetricNonHolomorphicMTransform}).
\end{description}


\subsubsection{Prove the conjectures}
\label{sss:ProveTheConjectures}

One should also prove the mentioned conjectures:
\begin{itemize}
\item The expressions for the external and internal radii, \smash{$R_{\textrm{ext}}$} and \smash{$R_{\textrm{int}}$}, of the mean spectral domain $\mathcal{D}$ for an arbitrary weighted sum $\mathbf{S}$ of the CUE's (end of paragraph~\ref{sss:EIGExample2}).
\item The ``$\erfc$ conjecture,'' concerning the validity of the form--factor (\ref{eq:FiniteSizeBorderlineFactor}), in the case of $\mathbf{S}$ with the CUE's (paragraph~\ref{sss:TheErfcFormFactorForWeightedSumsOfFreeUnitaryRandomMatrices}), and in full generality, \ie for any non--Hermitian model with rotationally--symmetric mean spectrum (paragraph~\ref{sss:TheErfcConjecture}).
\item The conjecture relating the mean densities of the eigenvalues and singular values, in the thermodynamic limit, when the former displays rotational symmetry around zero (\ref{eq:ConjectureAboutRotationallySymmetricSpectra}) (subsection~\ref{ss:ConjectureAboutRotationallySymmetricSpectra}).
\end{itemize}


\subsubsection{Other open problems}
\label{sss:OtherOpenProblems}

One might undertake the following research projects:
\begin{itemize}
\item Investigate, using the quaternion addition law (\ref{eq:QuaternionAdditionLawForS}), the weighted sum $\mathbf{S}$ for the \smash{$\mathbf{U}_{l}$}'s having more complicated probability distributions than CUE. In this case, generically, the mean spectral density will not be rotationally--symmetric around zero, hence, the conjecture (\ref{eq:ConjectureAboutRotationallySymmetricSpectra}) will not hold, and other means to reach the singular values will have to be devised.
\item A major research program would be to search for the full JPDF of the eigenvalues of $\mathbf{S}$ (and of $\mathbf{W}$), for arbitrary $L$, $K$, weights, as well as --- perhaps --- other probability distributions of the \smash{$\mathbf{U}_{l}$}'s and \smash{$\mathbf{A}_{k}$}'s.
\end{itemize}

Some other questions worth considering:
\begin{itemize}
\item Calculate the ``entanglement entropy'' (see \eg~\cite{SEMZyczkowski2010,SommersZyczkowski2004}), \smash{$S_{\mathbf{X}} \equiv - \int \dd x \rho_{\mathbf{X}^{\dagger} \mathbf{X}} ( x ) x \log x$}, for $\mathbf{X} = \mathbf{S}$ and $\mathbf{X} = \mathbf{W}$.
\item Analyze more thoroughly the behavior of the mean singular values density \smash{$\rho_{\mathbf{X}^{\dagger} \mathbf{X}} ( x )$}, for $\mathbf{X} = \mathbf{S}$ and $\mathbf{X} = \mathbf{W}$, near the end--points of the support (\emph{cf.} Remark 3 in paragraph~\ref{sss:SVTheMasterEquation}).
\end{itemize}


\begin{acknowledgments}
I am grateful to Piotr Bo\.{z}ek and Zdzis{\l}aw Burda for many stimulating discussions. I am grateful to Karol \.{Z}yczkowski for informing me about his recent work and inspiring me to investigate the model $\mathbf{W}$. I thank Eugene Kanzieper and Tim Rogers for valuable comments.

My work has been partially supported by the Polish Ministry of Science and Higher Education Grant ``Iuventus Plus'' No.~0148/H03/2010/70. I acknowledge the financial support of Clico Ltd., Oleandry 2, 30--063 Krak\'{o}w, Poland, while completing parts of this paper.
\end{acknowledgments}



\end{document}